\documentclass{article}
\usepackage[a4paper,
            bindingoffset=0.2in,
            left=0.6in,
            right=0.6in,
            top=1in,
            bottom=1in,
            footskip=.25in]{geometry}
\usepackage[utf8]{inputenc}
\usepackage{jcappub}
\usepackage{natbib}
\usepackage{fancyhdr}
\setcitestyle{square,numbers}
\usepackage{graphicx}
\usepackage{xcolor}
\usepackage{float}
\usepackage{multirow,array,longtable}
\usepackage{booktabs}
\usepackage{arydshln} 
\usepackage[makeroom]{cancel}
\usepackage{amsfonts}
\usepackage{slashed}

\usepackage{tikz-feynman}
\usepackage{subcaption}
\usepackage{tikz}
\usepackage{amsmath}
\tikzset{
    ->-/.style={decoration={
  markings,
  mark=at position .5 with {\arrow{>}}},postaction={decorate}},
    -<-/.style={decoration={
  markings,
  mark=at position .5 with {\arrow{<}}},postaction={decorate}},
    ->/.style={decoration={
  markings,
  mark=at position .4 with {\arrow{>}}},postaction={decorate}},
}
\usepackage{amssymb}
\usepackage{longtable}
 \usepackage{lscape}

\renewcommand\eqref[1]{Eq.~(\ref{#1})}

\newcommand\figref[1]{Fig.~\ref{#1}}

\newcommand\tabref[1]{Table~\ref{#1}}

\newcommand\secref[1]{Section~\ref{#1}}
\newcommand\appref[1]{Appendix~\ref{#1}}
\newcommand{\be}{\begin{equation}}
\newcommand{\ee}{\end{equation}}
\newcommand{\bear}{\begin{eqnarray}}
\newcommand{\eear}{\end{eqnarray}}
\newcommand{\nn}{\nonumber}

\def\1loop{one-loop}

\newcommand{\mL}{\mathcal{L}}

\newcommand{\mV}{\mathcal{V}}

\def\mh{m_{H}}
\def\mw{m_{W}}
\def\mz{m_{Z}}

\def\vev{{\it v}}

\title{Matching HEFT and SMEFT in double and triple Higgs production from weak boson fusion}
\author[a]{D.  Domenech,}
\author[a]{M.  Herrero,}
\author[b]{R. A. Morales}
\author[c]{and A. Salas-Bernárdez,}
\affiliation[a]{Departamento de F\'{\i}sica Te\'orica and Instituto de F\'{\i}sica Te\'orica, IFT-UAM/CSIC,\\
Universidad Aut\'onoma de Madrid, Cantoblanco, 28049 Madrid, Spain}
\affiliation[b]{IFLP, CONICET - Dpto. de F\'{\i}sica, Universidad Nacional de La Plata, \\ 
C.C. 67, 1900 La Plata, Argentina}
\affiliation[c]{Departamento de Análisis Matemático y Matemática Aplicada and IPARCOS, \\ 
Universidad Complutense de Madrid,  Plaza de las Ciencias 3, 28040 Madrid, Spain}

\emailAdd{jose.domenech@estudiante.uam.es}
\emailAdd{maria.herrero@uam.es}
\emailAdd{roberto.morales@fisica.unlp.edu.ar}
\emailAdd{alexsala@ucm.es}

\abstract{
In this work we study the matching between the two most popular effective field theories for beyond standard model Higgs physics,  SMEFT and HEFT.  To perform this matching we follow the approach of identifying the corresponding scattering amplitudes for physical processes in both theories,  instead of the most usual approach of relating the corresponding effective Lagrangians or effective actions.  In this work we focus  on the physical processes of double and triple Higgs production from weak boson fusion,  in particular,  $WW \to HH$, $ZZ \to HH$, $WW \to HHH$ and $ZZ \to HHH$ and complement them with the elastic scattering $HH \to HH$.   We present here the analytical solution to this matching in terms of relations among the coefficients in both theories and comment on the most relevant phenomenological implications of such relations for collider physics. }

\allowdisplaybreaks

\begin{document}
\begin{flushright}
	IFT-UAM/CSIC-25-52\\
	IPARCOS-UCM-25-030 
\end{flushright}
\maketitle
\section{Introduction}
\label{intro}
Since the discovery of the Higgs boson \cite{ATLAS:2012yve,CMS:2012qbp} there has been no clear evidence of new physics beyond the Standard Model (SM) of Particle Physics.  In the absence of any experimental signal of new fundamental particles,  resonances,  interactions,  nor distortions from SM interactions,  the most appropriate tool for studying Beyond Standard Model (BSM) physics is mostly agreed to be  provided by the Effective Field Theory (EFT) approach.  In this work we will focus on the EFT approach to describe BSM Higgs physics and will assume that the potential new physics appears exclusively in the bosonic sector,  including both the scalar and the gauge boson sectors.  Consequently,  the fermions and their couplings to the other SM particles are assumed here to be as in the SM.  As for the particular EFTs,  we will consider here the two most popular ones,  
nowadays widely employed for tests of BSM Higgs physics at colliders: the so-called HEFT (Higgs Effective Field Theory) and the SMEFT (Standard Model Effective Field Theory).  The advantage of using EFTs is that they allow for a test of the BSM Higgs physics  in a model independent way, i.e without assuming a particular underlying fundamental theory which operates at the ultraviolet (UV) high energy scale,  this being generically called $\Lambda$.  At low energies compared to $\Lambda$,  the information of the new Higgs physics  is encoded in an effective Lagrangian in terms of a  set of effective operators,  these being built with the SM fields (here the bosonic fields) and with the unique requirement of being invariant under the SM gauge symmetry,  $SU(3) \times SU(2) \times U(1)$.  The coefficients in front of these operators (usually called Wilson coefficients) are generically  unknown and  encode information of the particular underlying fundamental theory
(for reviews,  see for instance \cite{Brivio:2017vri,Dobado:2019fxe}).  For our study here,  we focus on the electroweak (EW) part of these two EFTs and restrict ourselves to the CP preserving effective operators. 

 The values of the EFT coefficients can  then be computed by integrating out the heavy modes from the particular assumed UV theory.   It is well known that depending on the kind of dynamics involved in the fundamental theory,  it is more appropriate the use of one EFT or another.  Usually, the SMEFT is more appropriate to describe the low energy behaviour of weakly interacting dynamics,  where by construction the heavy modes of the UV theory decouple from low energy physics and the observable effects left at the colliders go as inverse powers of $\Lambda$.  In contrast,  the HEFT is more appropriate to describe strongly interacting underlying UV dynamics where the heavy modes may not decouple from the low energy physics and the effects left at the colliders might not all go as inverse powers of $\Lambda$.  An interesting example of this different behaviour of the two EFTs is provided when considering  the particular case of the Two Higgs Doublet Model (2HDM) as the UV theory operating at the high energy $\Lambda$.  It has been shown that,  depending on the particular values assumed on the input 2HDM parameters (basically,  the masses $m_{h^0}$,  $m_{H^0}$,  $m_{H^{\pm}}$,  $m_{A^0}$,  and the mixing angles,  $\cos(\beta-\alpha)$ and $\tan \beta$)  and after integrating out the heavy scalars,   $H^0$, $H^\pm$ and $A^0$,  the resulting EFT with the lightest Higgs boson $h^0$ being identified with the observed Higgs particle $H$ of mass $m_H=125$ GeV,  can be either the SMEFT  \cite{Dawson:2023ebe} (with coefficients indicating decoupling  of $H^0$, $H^\pm$ and $A^0$),   or the HEFT \cite{Arco:2023sac} (with coefficients indicating non-decoupling  of $H^0$, $H^\pm$ and $A^0$).  The final aim in order to disentangled the two approaches is then to test with the present and future colliders which are the values of their respective EFT coefficients that fit best to experimental data.  

At the practical level,  the main differences between these two bosonic EFTs are,  on the one hand,   the representation used for the bosonic fields and,  on the other hand,  the criteria to order the effective operators built with these bosonic fields.  In the SMEFT the scalar fields (both the Higgs and the would be Goldstone bosons) are assumed to be in a linear representation of the electroweak symmetry group,  specifically in a doublet of $SU(2)$, and  the criterion to order the SMEFT effective operators is by increasing their canonical dimension.   In the HEFT,  the Higgs boson is a singlet under all symmetries (including $SU(2)$) whereas the would be Goldstone bosons are placed  in a non-linear representation of $SU(2)$,  usually by an exponential function.  The HEFT effective operators are ordered differently,  by their chiral dimension which counts 
the powers of momenta and masses ($m_H$,  $m_W$ and $m_Z$ ,  in the present case).  Thus,  one usually refers to truncations of the respective Lagrangians as: 1) NLO-SMEFT which includes up to canonical dimension 6 effective operators and  2) NLO-HEFT which includes up to chiral dimension 4 effective operators.  On the other hand,  both EFTs incorporate the gauge particles by the gauge principle based on the same gauge symmetry of the SM, i.e.  $SU(3) \times SU(2) \times U(1)$.  Both SMEFT and HEFT are renormalizable in the weaker sense of EFTs,  and the one-loop renormalization in the generic $R_\xi$ covariant gauges has been provided in the literature for both,  the SMEFT \cite{Dedes:2017zog} and the HEFT \cite{Herrero:2020dtv,  Herrero:2021iqt,  Herrero:2022krh}.  For more details and relevant effective operators regarding multiple Higgs boson production within HEFT,  see also \cite{Contino:2010mh,Contino:2013gna,Anisha:2024ljc, Anisha:2024ryj}. 

The question of how to relate both theories has also been considered  in the literature.  Most frequently these two EFTs  are related by comparing their respective truncated Lagrangians.  For instance,  the NLO-HEFT and the NLO-SMEFT can be related by comparing their respective predictions  for the so-called anomalous couplings ($\kappa_W$,  $\kappa_Z$, 
$\kappa_\gamma$,  etc.). This is equivalent to compare their respective predictions for the interaction vertices,  i.e.  their corresponding Feynman rules in the momentum space.  This is done,  for instance,  in ref. \cite{Brivio_2014},  where they conclude on the role of the so-called SMEFT at dimension six siblings effective operators and also on the decorrelation/correlation among the effective coefficients when comparing HEFT and SMEFT (see also,  \cite{Gavela:2016vte}).  The use of coupling modifiers $\kappa_V$,  $\kappa_{2V}$ , etc.,   to disentangle  BSM patterns from the two SMEFT and HEFT approaches at colliders has also been considered in the literature,  see for instance \cite{Domenech:2022uud, Englert:2023uug}.  Another type of comparison of the HEFT and the SMEFT at the Lagrangian level has been done,  in this case  for the pure scalar theory (Higgs and GBs),  in refs. \cite{Salas-Bernardez:2022hqv,Gomez-Ambrosio:2022qsi,  Gomez-Ambrosio:2022why,  Delgado:2023ynh}. 
They relate the truncated Lagrangians of HEFT and  SMEFT by performing a series of transformations of the scalar fields and arrive to particular relations among the HEFT  and the  SMEFT coefficients that are involved in the scalar field interactions.   Another different and more complete way to compare these two theories is by their respective effective actions, or equivalently their predictions for the complete set of 
1PI (one-particle irreducible) functions (see,  for instance,  the review \cite{Brivio:2017vri} and references therein).   
Another distinct techniques apply on-shell methods in EFTs to calculate tree level amplitudes with the spinor helicity formalism. These techniques have been used in the literature in the recent years to construct the operator basis, both in the HEFT~\cite{Sun:2022ssa,Dong:2022jru} and in the SMEFT~\cite{Goldberg:2024eot}. They have also been applied to compare the two EFTs, HEFT and SMEFT, in some particular cases of two-to-two scattering on-shell amplitudes (see, for instance,  ref.~\cite{Liu:2023jbq}). 
Finally, there are also  comparisons using arguments based on a geometric description of the scalar field space \cite{Alonso:2015fsp,  Alonso:2016btr, Alonso:2016oah, Cohen:2020xca, Alonso:2021rac}.  Within this geometric approach,  it is 
well established that SMEFT can always be cast in HEFT form and not the other way around,  therefore,   it is concluded that HEFT is more general than SMEFT. 

In the present work,  we proceed as in ref.~\cite{Domenech:2022uud}  by comparing these two EFTs  in a  different way,   which we believe is better motivated for phenomenological purposes.  Instead of comparing their Lagrangian and/or interaction vertices,  we compare their respective predictions for scattering and decay amplitudes which are the mathematical objects directly related to physical observables,  i.e.  to cross sections,  differential cross sections,  total and partial decay widths,  etc.  More specifically,  we choose a set of physical processes,   make analytical predictions for the corresponding amplitudes of these processes from the two truncated NLO-HEFT and NLO-SMEFT at the tree level,   identify correspondingly  these amplitudes of the two theories,  and finally solve analytically for the solution of the coefficients of the two EFTs fulfilling these matching equations for the full set of amplitudes considered.   As it will be stated here,  the predicted amplitudes in both theories are Lorentz and gauge invariant functions,  and therefore the solutions for the coefficients from the matching equations are  unambiguously determined.  The set of processes considered here are motivated by present and future searches of new Higgs physics in multiple Higgs production at colliders.  In particular,  we include here the following electroweak processes:  single,  double and triple Higgs production from Weak Boson Fusion (WBF),  $WW \to H$,  $ZZ \to H$,  $WW \to HH$,  $ZZ \to HH$,  $WW \to HHH$,  $ZZ  \to HHH$,  and add,  for completeness,  the case $HH \to HH$.  Then we solve in this paper the HEFT/SMEFT  matching equations for all the corresponding physical amplitudes,  including all possible polarizations for the external gauge bosons (longitudinal and transverse).
Finally,  we conclude by presenting the solutions to all these matching equations in terms of the relations among their corresponding HEFT/SMEFT  coefficients and comment on the potential implications for collider phenomenology  of those relations. 
Particular attention will be devoted to the contact interactions $WWHHH$ and $ZZHHH$, which naturally arise at LO in the HEFT. On the experimental side, a recent ATLAS analysis \cite{ATLAS:2024xcs} focuses on triple Higgs boson production in the 6 $b$-jets final state, as proposed in \cite{Papaefstathiou:2019ofh,Papaefstathiou:2023uum} (see also \cite{Abouabid:2024gms,ATL-PHYS-PUB-2025-003} for the key challenges in future $HHH$ searches).

The paper is organized as follows.   In \secref{HEFToperators} we review the relevant NLO-HEFT bosonic operators for the processes of our most interest here,  multiple  Higgs production from Weak Boson Fusion.   In \secref{SMEFToperators} we review  the NLO-SMEFT relevant bosonic operators for these same processes.  In  \secref{amplitudes} we present the analytical predictions for the scattering amplitudes of those processes from both NLO-HEFT and NLO-SMEFT in terms of the selected effective bosonic operators and the corresponding EFTs coefficients.  In \secref{matching} we discuss the matching equations for the amplitudes of the selected processes and present their solutions in terms of relations among the corresponding HEFT and SMEFT coefficients.  Finally, in \secref{conclu} we summarize the conclusions and comment on the phenomenological implications at colliders.

\section{Relevant HEFT operators for double and triple Higgs WBF production}
\label{HEFToperators}
The relevant EW effective bosonic operators for the present work are contained in the so-called Next to Leading Order (NLO) HEFT Lagrangian (also called Electroweak Chiral Lagrangian in the literature).   Here,  we restrict ourselves to the CP invariant subset of effective operators.  It is customary to present this NLO Lagrangian separated into two pieces,  following the standard conventions for Chiral Lagrangians,  where the first piece is assigned to chiral dimension 2,  usually named leading order (LO) Lagrangian or ${\mL}_2$,  and the second piece ${\mL}_4$ is assigned to chiral dimension 4.  This separation in chiral dimension is not really necessary for the present work,  but helps in understanding the posterior comparison of the corresponding amplitudes for the HEFT and the SMEFT.  Remember that in the Chiral Lagrangian context it is customary to assign a double role to the ${\mL}_4$ part.  On one hand,  it provides the needed counterterms  for the renormalization of the one-loop contributions from ${\mL}_2$.  On the other hand,  it provides new contributions to the scattering amplitudes that encode  BSM physics,   which can be checked experimentally.  These contributions from ${\mL}_4$ in the scattering amplitudes  are considered in momentum space as NLO respect to the LO contributions from ${\mL}_2$,  since they count as ${\cal O}(p^4)$,  where $p$ refers to either momentum or mass,  in contrast to the contributions from ${\mL}_2$,   that count as ${\cal O}(p^2)$.  For the present work we will compute all the scattering amplitudes at the tree level and,  therefore,  the role played by  the ${\mL}_4$ part will be just the second one.  

We summarize the HEFT Lagrangian that is relevant for this work in the following:   
  \begin{eqnarray}
   \label{eqn:NLO-HEFT}
  {\mL}^{\rm NLO}_{\rm HEFT}  \, = \, {\mL}_2 +{\mL}_4
    \end{eqnarray}
The LO Lagrangian  is given by:
     \begin{eqnarray}
      {\mL}_2   &=& \frac{v^2}{4}  \left(1 + 2  \textcolor{red}{a} \frac{H}{v} +\textcolor{red}{b} \frac{H^2}{v^2}+ \textcolor{red}{c}  \frac{H^3}{v^3} \right)
        \text{Tr}[D_{\mu} U^{\dagger} D^{\mu} U] + \frac12 \partial_{\mu} H \partial^{\mu} H - V(H) \nn \\
       && - \frac{1}{2 g^2} \text{Tr}[\hat{W}_{\mu \nu} \hat{W}^{\mu \nu}] - \frac{1}{2 g'^2} \text{Tr}[\hat{B}_{\mu \nu} \hat{B}^{\mu \nu}] + {\mL}_{GF} + {\mL}_{FP},
        \label{eqn: L2}
    \end{eqnarray}
The relevant fields and quantities appearing in this Lagrangian are:  
$H$ is the Higgs field which is introduced in the HEFT as a singlet field, in contrast to the SM or the SMEFT where it is introduced as a component of the usual doublet 
$\Phi$.  Due to this singlet Nature of $H$ within the HEFT,  the first term in ${\mL}_2$ contains a polynomial function in powers of $(H/v)$ which for the present work we take up to cubic powers.  Then,  there are three coefficients involved in this function,  $a$, $b$ and $c$.  The three of them will have consequences for the phenomenology of multiple Higgs production from WBF that we will discuss later.

The $U$ field is a $2\times2$ matrix
\begin{equation}
        U \, = \, \exp \left( i \frac{{\omega_i} {\tau_i}}{v} \right),
    \end{equation}  
   that contains the three GB fields ${\omega}_i$ ($i=1,2,3$) in a non-linear representation of the $SU(2)$ symmetry group and 
   $\tau_i$ ($i=1,2,3$) are the three Pauli matrices.

   The EW covariant derivative of this $U$ field 
   is defined as:
   \begin{equation}
        D_{\mu} U \, = \, \partial_{\mu} U - i \hat{W}_{\mu} U + i U \hat{B}_{\mu},
    \end{equation} 
    that contains the EW gauge fields,  $\hat{W}_{\mu} = \frac{g}{2} W^i_{\mu} \tau^i$ and $\hat{B}_{\mu} = \frac{g'}{2} B_{\mu} \tau^3$ and the EW gauge couplings $g$ and $g'$.  The corresponding EW field strength tensors are given by:
       \begin{equation}
        \hat{W}_{\mu \nu} \, = \, \partial_{\mu} \hat{W}_{\nu} - \partial_{\nu} \hat{W}_{\mu} - i [\hat{W}_{\mu}, \hat{W}_{\nu}], \hspace{8mm} \hat{B}_{\mu \nu} \, = \, \partial_{\mu} \hat{B}_{\nu} - \partial_{\nu} \hat{B}_{\mu}.
    \end{equation}
    The physical gauge fields are then given,  as usual,  by:
\be
W_{\mu}^\pm = \frac{1}{\sqrt{2}}(W_{\mu}^1 \mp i W_{\mu}^2) \,,\quad
Z_{\mu} = c_W W_{\mu}^3 - s_W B_{\mu} \,,\quad
A_{\mu} = s_W W_{\mu}^3 + c_W B_{\mu} \,,
\label{eq-gaugetophys}
\ee
where we use the short notation $s_W=\sin \theta_W$ and $c_W=\cos \theta_W$,  with $\theta_W$ the weak angle.
$V(H)$ is the Higgs potential within the HEFT, which includes the triple and quartic  Higgs self-interactions:
    \begin{equation}
        V(H) \, = \, \frac12 m_H^2 H^2 + \textcolor{red}{\kappa_3} \lambda v H^3 + \textcolor{red}{\kappa_4} \frac{\lambda}{4} H^4.
    \label{VHEFT}    
    \end{equation}
The relations among couplings  and masses within the LO-HEFT are:   $m_H^2  = 2 \lambda v^2$,  $\mw=g v /2$ and $\mz=\mw/c_W$.   The vacuum expectation value  $v$ can be written  in terms of  the Fermi coupling constant $G_F$ by $v=(\frac{1}{\sqrt{2} G_F})^{1/2} =246$ GeV. 

Finally,  ${\mL}_{GF}$ and ${\mL}_{FP}$ are the gauge fixing  and Faddeev-Popov terms,  respectively, whose explicit expressions in the general $R_\xi$ covariant gauges can be found in 
\cite{Herrero:2020dtv, Herrero:2021iqt,Herrero:2022krh},  together with more definitions and specifications within the HEFT (see also,  \cite{Anisha:2024ljc, Anisha:2024ryj}).

The  remaining operators in the  NLO Lagrangian that are relevant for the present computation of double and triple Higgs production from WBF are contained in ${\mL}_4$.  We mention here only those affecting the interactions of the scalar sector with the EW gauge bosons and the self-interactions of the scalar sector 
\cite{Herrero:2022krh,  Anisha:2024ljc, Anisha:2024ryj} (for a more complete list of NLO effective operators see  \cite{Brivio_2014}):
 \begin{eqnarray}
{\mL}_{4} \, &=& -a_{dd\mV\mV 1} \frac{\partial^\mu H\,\partial^\nu H}{v^2} {\rm Tr}\Big[ {\cal V}_\mu {\cal V}_\nu \Big] -a_{dd\mV\mV 2} \frac{\partial^\mu H\,\partial_\mu H}{v^2} {\rm Tr}\Big[ {\cal V}^\nu {\cal V}_\nu \Big]  \nn\\
&& -\frac{\mh^2}{4}\left(2a_{H\mV\mV}\frac{H}{v}+a_{HH\mV\mV}\frac{H^2}{v^2}\right) {\rm Tr}\Big[ {\cal V}^\mu {\cal V}_\mu \Big]  \nn\\
&& +\left( \textcolor{red}{a_{H0}} \frac{H}{v}+ \textcolor{red}{a_{HH0}}\frac{H^2}{v^2}+
\textcolor{red}{a_{HHH0}}\frac{H^3}{v^3} \right ) \left( \mz^2-\mw^2 \right) {\rm Tr}\Big[ U \tau^3 U^\dagger {\cal V}_\mu \Big] {\rm Tr}\Big[ U \tau^3 U^\dagger {\cal V}^\mu \Big] 
\nn\\
&& +\left(\hat a_{Hdd} \frac{H}{\vev} + \hat a_{HHdd} \frac{H^2}{v^2} \right) \partial^\mu H\,\partial_\mu H + a_{dddd} \frac{1}{v^4}  \partial_\mu H \partial^\mu H \,  \partial_\nu H \partial^\nu H \nn\\
&& -\left(\textcolor{red}{a_{HWW}} \frac{H}{\vev} +\textcolor{red}{a_{HHWW}} \frac{H^2}{v^2}\right) {\rm Tr}\Big[\hat{W}_{\mu\nu} \hat{W}^{\mu\nu}\Big] +i\left(a_{d2} +a_{Hd2}\frac{H}{v}\right)\frac{\partial^\nu H}{v} {\rm Tr}\Big[ \hat{W}_{\mu\nu} {\cal V}^\mu\Big]  \nn\\
&& -\left(\textcolor{red}{a_{HBB}} \frac{H}{\vev} +\textcolor{red}{a_{HHBB}} \frac{H^2}{v^2}\right) {\rm Tr}\Big[\hat{B}_{\mu\nu} \hat{B}^{\mu\nu}\Big] +i\left(a_{d1} +a_{Hd1}\frac{H}{v}\right)\frac{\partial^\nu H}{v} {\rm Tr}\Big[ U\hat{B}_{\mu\nu}U^\dagger {\cal V}^\mu\Big]  \nn\\
&& +\left(\textcolor{red}{a_{H1}} \frac{H}{\vev} +\textcolor{red}{a_{HH1}} \frac{H^2}{v^2}\right) {\rm Tr}\Big[U\hat{B}_{\mu\nu}U^\dagger \hat{W}^{\mu\nu}\Big]
\label{eq-L4}
\end{eqnarray}   
where we have used the usual definition ${\cal V}_\mu \equiv  (D_\mu U)U^\dagger$ and a short notation with respect to \cite{Herrero:2022krh,Anisha:2024ljc} given by 
$\hat a_{Hdd}=a_{Hdd} \frac{m_H^2}{v^2}+a_{ddW} \frac{m_W^2}{v^2}+ a_{ddZ } \frac{m_Z^2}{v^2}$ and $\hat a_{HHdd}=a_{HHdd} \frac{m_H^2}{v^2}+a_{HddW} \frac{m_W^2}{v^2}+ a_{HddZ } \frac{m_Z^2}{v^2}$.
Notice that in the list above,  the full set of bosonic operators in the NLO Lagrangian,  displayed in ref. \cite{Brivio_2014},   has been reduced to a minimal set by the use of the equations of motion (for further details,  and the relation among the various notations,  see  \cite{Herrero:2022krh}).  

It is important to remark that the previous HEFT Lagrangian in \eqref{eqn:NLO-HEFT} is invariant under the EW  $SU(2)_L \times U(1)_Y$ gauge symmetry and is renormalizable in the weaker sense of EFTs with a truncated Lagrangian.  The one-loop renormalization of the previous truncated  HEFT Lagrangian in \eqref{eqn:NLO-HEFT} is possible by the proper renormalization of the HEFT parameters involved,  including the particle masses,  couplings,  fields and the HEFT coefficients,  both LO ($a$,  $b$,  $c$,  $\kappa_3$, $\kappa_4$)  and NLO ones ($a_i$'s) .  The one-loop renormalization of the bosonic HEFT in $R_\xi$ covariant gauges has been performed in a series of works,  see refs. \cite{Herrero:2020dtv, Herrero:2021iqt,Herrero:2022krh, Anisha:2024ljc},  where the counterterms of the HEFT parameters and fields in \eqref{eqn:NLO-HEFT} have been set. 

\begin{figure}[!t]
    \centering
        \includegraphics[scale=0.3]{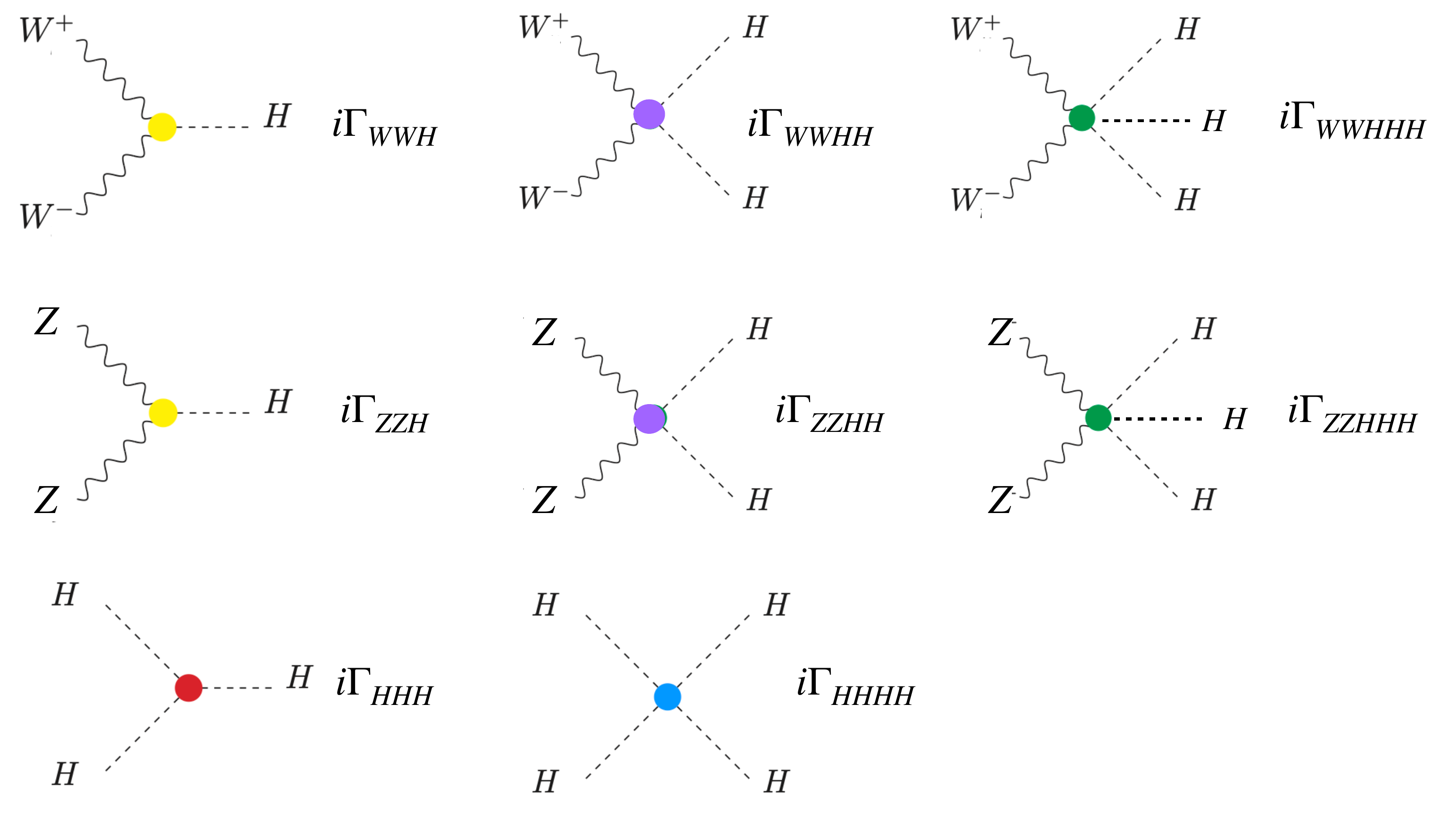}
    \caption{Relevant effective vertices for double and triple Higgs production.  The corresponding values of $\Gamma_{WWH}$,  $\Gamma_{WWHH}$,  
    $\Gamma_{ZZH}$,  $\Gamma_{ZZHH}$,  $\Gamma_{HHH}$ and $\Gamma_{HHHH}$ are given in \appref{apHEFT} for HEFT and in \appref{apSMEFT} for SMEFT.  We use different colors in these BSM vertices to visually differentiate them more clearly.  For comparison,  the corresponding SM Feynman rules are also included in \appref{apSM}.}
    \label{FR}
    \end{figure}

The Feynman rules of the relevant HEFT vertices  are summarized in \figref{FR} and in \appref{apHEFT}.  These relevant vertices include the effective interactions of the Higgs boson $H$ with the electroweak gauge bosons $V=W^{\pm}, Z$ and the Higgs self-interactions.  More specifically we focus on the following interactions:  
$VVH$,  $VVHH$,  $VVHHH$,   with $VV=WW,ZZ$,  and $HHH$,  $HHHH$.  As it can be seen  in these Feynman rules, both  the LO  and NLO HEFT coefficients are involved in these effective interactions. 
Moreover,  as it will be shown in the forthcoming sections,  we wish to remark that  only a subset of these HEFT coefficients  will be needed to solve the matching of the HEFT and SMEFT amplitudes that we are aiming to provide in this work.   Specifically,  we will present analytical results for the scattering amplitudes of the following specific processes including single,  double and triple Higgs production: $WW \to H$,  $ZZ \to H$,   $HH \to HH$,  $WW \to HH$,  $ZZ\to HH$,  $WW \to HHH$ and $ZZ \to HHH$.  These processes have been selected because they participate as possible subprocesses in multiple Higgs production at future lepton and hadron colliders.  In particular those subprocesses with initial EW gauge boson pairs are essential for the production mechanism called  single $H$,  double $HH$ and triple $HHH$ production from Weak Boson Fusion.   

For our main objective here,  i.e.  for solving the matching HEFT/SMEFT,  we have selected the proper subset of HEFT coefficients which are marked in red color in the HEFT Lagrangian.  It will be shown later that these HEFT coefficients being involved in the selected processes are the following 14 specific coefficients: $a$, $b$, $c$, $\kappa_3$,  $\kappa_4$,  $a_{H0}$,  $a_{HH0}$,  $a_{HHH0}$,  $a_{HWW}$,  $a_{HHWW}$, $a_{HBB}$,  $a_{HHBB}$,  $a_{H1}$ and $a_{HH1}$.  

Regarding the comparison of the HEFT with the SM,  it can also be checked that the SM Feynman rules, summarized in \appref{apSM},  can be reached from the HEFT Feynman rules,  summarized  in \appref{apHEFT},  by setting $a=b=\kappa_3=\kappa_4=1$,  $c=0$ and $a_i=0$ for the rest of HEFT coefficients.  This is equivalent to say that the proper choice for  the HEFT parameters that match the SM predictions for the scattering amplitudes are: $a=b=\kappa_3=\kappa_4=1$ and $c=0$.  The rest of HEFT coefficients $a_i$ in ${\mL}_4$ must also be set to zero in order to match the HEFT with the SM.  Notice that the contact interactions parametrized by $c$ are absent at the tree level within the SM.  Thus,  any experimentally detectable effect from $c \neq 0$ and/or $a_i \neq 0$ will provide clear signals of BSM Higgs Physics at colliders \cite{Gonzalez-Lopez:2020lpd,Domenech:2022uud,  Davila:2023fkk,  Anisha:2024ljc, Anisha:2024ryj}. 

Regarding the codes used in this work,  for the forthcoming computations we use Wolfram Mathematica \cite{WMathematica} with our own  model file that implements the previous HEFT Lagrangian in the $R_\xi$ gauges and derive the Feynman rules and the scattering amplitudes  with the help of FeynRules \cite{FeynRules} and FormCalc \cite{FormCalc-LT}.  We have also performed  cross checks with the help of FeynCalc \cite{FeynCalc}.  As for the drawing of Feynman diagrams we use FeynArts \cite{FeynArts}.

For phenomenological discussions it is convenient to relate these LO-HEFT coefficients  with the so-called $\kappa$ modifiers which parametrize the coupling deviations with  respect to the SM couplings.   In particular: 
\begin{equation}
\label{kappas}
a=\kappa_V \,\,\,, \, \, \, b=\kappa_{2V} \,\, \,, \,\,\,\kappa_3=\kappa_\lambda .
\end{equation}
Some progress for testing these anomalous couplings  by multiple Higgs production from Vector Boson Fusion at LHC is being discussed in the literature,  with particular focus on improving the precision tools  by using the $\kappa$-formalism (see, for instance,~\cite{Braun:2025hvr,Jager:2025isz}).

Generically,  as said above,  the LO-HEFT parameters $a$,  $b$,  $\kappa_3$ and $\kappa_4$ are set to 1 for the matching to the SM.  Then,  the deviations of the LO-HEFT predictions from the SM predictions should be parametrized  in terms of the differences of these LO-HEFT coefficients with respect to 1.   We define here these differences as:  $\Delta a \equiv 1-a $,  $\Delta b\equiv 1-b$,  $\Delta \kappa_3  \equiv 1-\kappa_3$,  and $\Delta  \kappa_4 \equiv 1-\kappa_4$. 

The present data from LHC set important constraints on these $\kappa$ parameters of \eqref{kappas},  which can be translated directly into constraints on the LO-HEFT parameters $a$,  $b$ and $\kappa_3$.  The quartic coupling $\kappa_4$ is not constrained yet.  These constraints can be shortly summarized 
as (see \cite{ATLAS:2022vkf, ATLAS:2024ish, ATLAS:2022jtk,CMS:2024awa,CMSPASHIG20011} for the corresponding analysis' assumptions):
\begin{equation}
\label{ATLASconstraints}
{\rm ATLAS} \,\,: \, \,\kappa_V \in  (0.96,1.11)  (95\% CL)\,\,,  \,\,\kappa_{2V} \in (0.6,1.5) (95\% CL) \,\,,  \,\, \kappa_\lambda \in (-0.4,6.3) (95\% CL)
\end{equation}
\begin{equation}
\label{CMSconstraints}
{\rm CMS} \,\,: \, \,\kappa_V \in  (0.97 , 1.09)  (95\% CL)\,\,,  \,\,\kappa_{2V} \in (0.6, 1.5) (95\% CL) \,\,,  \,\, \kappa_\lambda \in (-1.4,7.8) (95\% CL)
\end{equation}

\section{Relevant SMEFT operators for double and triple Higgs WBF production}
\label{SMEFToperators}
Within the SMEFT,  the set of  EW  bosonic effective operators  that are relevant for the present work are contained in the following NLO SMEFT Lagrangian.
 \begin{eqnarray}
   \label{eqn:NLO-SMEFT}
  {\mL}^{\rm NLO}_{\rm SMEFT}  \, = \, {\mL}^{(4)}_{\rm SMEFT} + {\mL}^{(6)}_{\rm SMEFT} 
    \end{eqnarray}
    where the LO-SMEFT Lagrangian,  with canonical dimension 4,  is the SM Lagrangian of the bosonic sector,  and ${\mL}^{(6)}_{\rm SMEFT}$ contains the set of relevant SMEFT operators with canonical dimension 6.  We use here the basis of operators and notation as defined in \cite{Grzadkowski_2010}.  
 The LO-SMEFT Lagrangian can then be written as:   
\bear 
\label{SMlag4}
 \, {\mL}^{(4)}_{\rm SMEFT}= {\mL}_{\rm SM}   &=&
                   -\frac{1}{4} W_{\mu\nu}^I W^{I\mu\nu}
                   -\frac{1}{4} B_{\mu\nu}   B^{\mu\nu} +  {\mL}_{GF}^{\rm SM} + {\mL}_{FP}^{\rm SM} \nn \\
&&   + \left( D_\mu \Phi \right)^\dagger \left( D^\mu \Phi \right) 
   + m^2 \Phi^\dagger \Phi -\frac{1}{2} \lambda \left( \Phi^\dagger \Phi \right)^2  ,
\eear
where $\Phi$ is the SM complex scalar doublet,  
\bear 
\Phi & =& \left( \begin{array}{c} \phi^+ \\ \frac{1}{\sqrt{2}} (v +
    H+ i \phi^0) \end{array} \right) \;, \quad
\label{doubletdef}
\eear 
and the conventions for the covariant derivative and SM electroweak field strengths are set here as in  \cite{Grzadkowski_2010}:
\bear
  D_\mu &=&  \partial_\mu+i g' B_\mu Y + i g W^{I}_\mu T^{I} \,\,, \, \, I= 1,2,3   \nn\\
  W_{\mu\nu}^I &=& \partial_\mu W^{I}_\mu-\partial_\nu W^I_\mu -g \epsilon^{IJK} W^J_\mu W^K_\nu \nn\\
   B_{\mu\nu} &=& \partial _\mu B_\nu -\partial_\nu B_\mu
\label{SMdefs}   
\eear
The definitions of the gauge fixing Lagrangian in the covariant $R_\xi$ gauges,   ${\mL}_{GF}^{\rm SM}, $ and the Fadeev-Popov Lagrangian,  ${\mL}_{FP}^{\rm SM}$,  are the usual ones in the SM.  It should be noticed that the usual normalization in the SMEFT for the Higgs self-coupling in \eqref{SMlag4} is different than its usual normalization in the HEFT in  \eqref{VHEFT}.  Specifically,  $\lambda_{\rm SMEFT} =2 \lambda_{\rm HEFT}$.  

The operators  of canonical dimension 6 that are relevant for the present work are:
\bear
{\mL}^{(6)}_{\rm SMEFT} \, &=& \frac{c_\Phi}{\Lambda^2} (\Phi^\dagger \Phi)^3 +\frac{c_{\Phi\Box}}{\Lambda^2}(\Phi^\dagger \Phi)\Box(\Phi^\dagger \Phi) +\frac{c_{\Phi D}}{\Lambda^2}(\Phi^\dagger D^\mu\Phi)^*(\Phi^\dagger D_\mu\Phi)  \nn\\
&& +\frac{c_{\Phi W}}{\Lambda^2}(\Phi^\dagger \Phi) W^I_{\mu\nu}W^{I\,\mu\nu} +\frac{c_{\Phi B}}{\Lambda^2}(\Phi^\dagger \Phi) B_{\mu\nu}B^{\mu\nu} +\frac{c_{\Phi WB}}{\Lambda^2}(\Phi^\dagger \tau^I\Phi) W^I_{\mu\nu}B^{\mu\nu}
\label{eq-LSMEFTdim6}
\eear
Notice again that we have restricted ourselves to the CP invariant subset of operators.  Then,  there are in total six Wilson coefficients involved: $c_\Phi$,  
$c_{\Phi\Box}$,  $c_{\Phi D}$,  $c_{\Phi W}$,  $c_{\Phi B}$,  and $c_{\Phi WB}$.  All these $c_i$'s  are dimensionless and we keep the prefactor $1/\Lambda^2$ explicitly in the forthcoming computation.   Again,  it is important to remark that the previous SMEFT Lagrangian in \eqref{eqn:NLO-SMEFT} is invariant under the electroweak $SU(2)_L \times U(1)_Y$ gauge symmetry.  The one-loop renormalization in $R_\xi$ covariant gauges of the previous SMEFT Lagrangian has been set,  and the full list of Feynman rules have also been provided in \cite{Dedes:2017zog}

The Feynman rules of the relevant SMEFT effective vertices  are summarized in \figref{FR} and in \appref{apSMEFT}.  For the following computations we use the rules from the SmeftFR-v3 package \cite{Dedes:2023zws}.  As in the previous HEFT case,  
the computations of the scattering amplitudes are performed with the help of FeynRules \cite{FeynRules} and FormCalc \cite{FormCalc-LT}.  We have also performed  cross checks with the help of FeynCalc \cite{FeynCalc}.  As for the drawing of Feynman diagrams we use FeynArts \cite{FeynArts}.


\section{HEFT and SMEFT  amplitudes for $VV \to HH$ and $VV \to HHH$}
\label{amplitudes}
\begin{figure}[!t]
\vspace{-2cm}   
    \centering
        \includegraphics[scale=0.4]{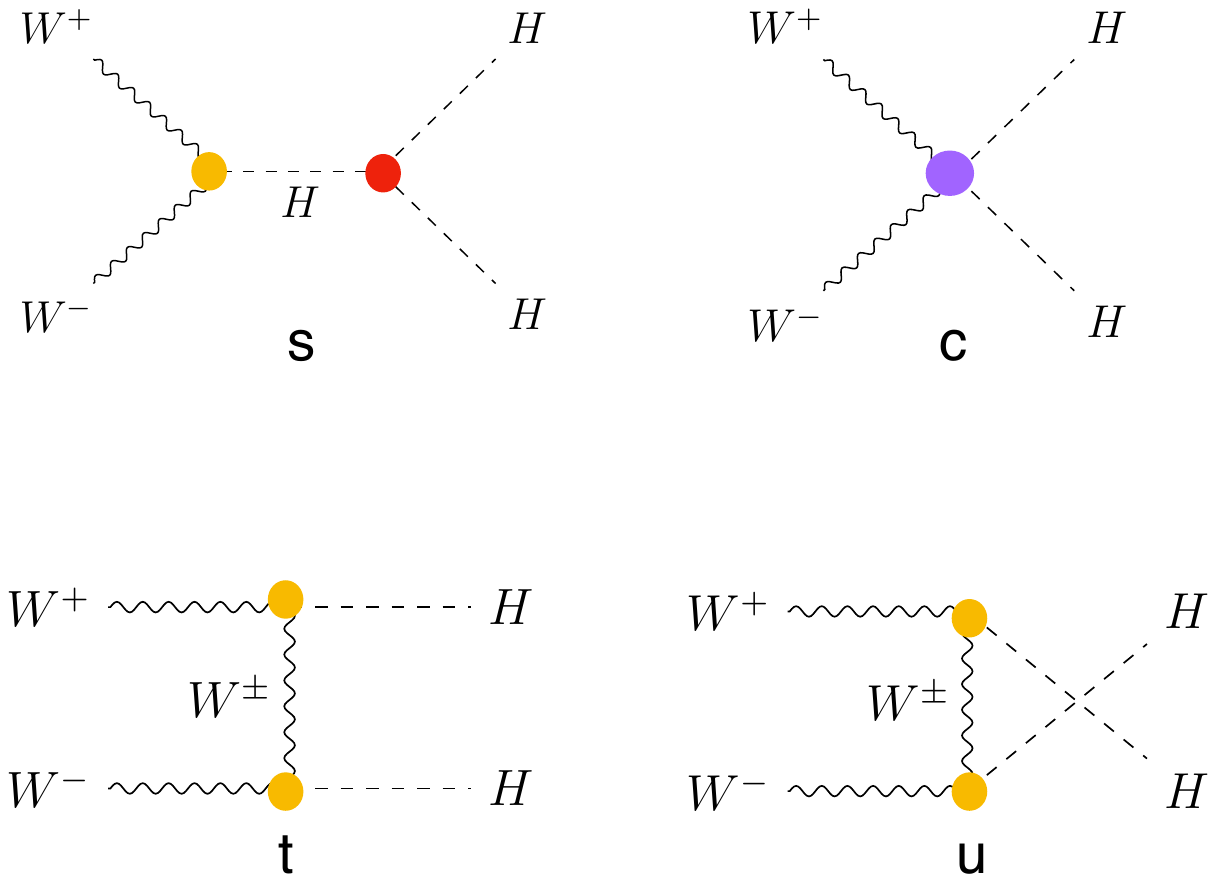}
    \caption{Feynman diagrams for $WW\to HH$ scattering in the unitary gauge.  The labels s, c, t and u refer to s-channel, c-channel, t-channel and u-channel respectively. The diagrams for $ZZ \to HH$ are similar but changing $W$ by $Z$.}
  \label{FDHH}  
    \end{figure}
Here we summarize the analytical results for double and triple Higgs production from WBF.  The computation is performed at the tree level.  We choose to compute  them in the unitary gauge and present here the results separately for  all the contributing diagrams.  The diagrams for $VV \to HH$ are displayed in \figref{FDHH}.   The diagrams for $VV \to HHH$ are displayed in \figref{FDHHH}.  

An important point that is worth to emphasize here is the explicit demonstration that all these scattering amplitudes are gauge invariant.  In particular,  we have  checked the full agreement  in the predictions of all the scattering amplitudes when computed in  the unitary gauge with the results when computed in the $R_\xi$  gauges.  Notice that in this later case and the full set of diagrams including also the internal Goldstone bosons (not shown here, for shortness) must be considered.   This check involves  
to demonstrate explicitly  the 
$\xi$  independence  of the resulting amplitude in all the studied cases for $VV \to HH$ and $VV \to HHH$ with $VV=WW$ and $VV=ZZ$ and for the two EFT's predictions,  HEFT and SMEFT.   This gauge invariance of all the scattering amplitudes is an important feature for the forthcoming comparison of the resulting HEFT and SMEFT amplitudes and to solve  the corresponding matching equations.  Besides,  and  
by  means of the  computation of these amplitudes within the SM in the $R_\xi$ gauges (not shown here for brevity),   we have also checked  the $\xi$ independence of all the SM amplitudes,  as expected.  

We summarize all these HEFT and SMEFT analytical  results in the following subsections.   We present all these results in the physical scheme where the amplitudes are written in terms of physical parameters for on-shell external particles,  namely,  the physical masses,  $m_H$, $m_W$, $m_Z$ and the vacuum expectation value related to the Fermi constant $v=(\frac{1}{\sqrt{2} G_F})^{1/2} =246$ GeV.  In addition,  the HEFT and SMEFT results are given in terms of the corresponding EFT coefficients.  Notice that the following results are for all the possible polarizations of the external EW gauge bosons, including both longitudinal and transverse modes.  Each particular helicity amplitude can be fixed by setting the proper polarization vectors in our analytical results.   Finally,  in the cases where the scattering is initiated by $ZZ$,  and for both HEFT and SMEFT results, we have used in addition the convenient short notation for some frequently appearing mass combinations:
\bear
c_W^2&=&\frac{m_W^2}{m_Z^2},  \, s_W^2= \frac{m_Z^2-m_W^2}{m_Z^2},  \,  s_W c_W= \frac{m_W \sqrt{m_Z^2-m_W^2}}{m_Z^2} .
\label{short}
\eear

\subsection{$ WW \to H H$}
We use the following momenta assignments $ W^+(k_1)W^-(k_2) \to H(k_3) H(k_4)$ with $k_1$ and $k_2$ ingoing and $k_3$ and $k_4$ outgoing.  The conservation of momentum is, then,  $k_1+k_2=k_3+k_4$.  To simplify we  also used here,  $\epsilon_1 \cdot k_1=0$, $\epsilon_2 \cdot k_2=0$,  where $\epsilon_1$ and $\epsilon_2$ are the polarization vectors of 
the ingoing $W^+$ and $W^-$ respectively.  

\subsubsection{HEFT}
The HEFT results for the scattering amplitude $A(WW \to HH)$ contain two parts  corresponding to the contributions from ${\cal L}_2$  and ${\cal L}_4$ respectively.  We present them here separated by channels.  The diagrams for these channels are  shown in \figref{FDHH}.  The part from ${\cal L}_2$ coincides with the SM result if one sets $a=b=\kappa_3=1$.  The part from ${\cal L}_4$ contains the $a_i$'s HEFT coefficients.  For brevity, we keep here  just the subset of HEFT coefficients that are needed to solve the forthcoming matching of the HEFT and the SMEFT amplitudes.  For this amplitude,  these are $a$,  $b$,  $\kappa_3$,  $a_{HWW}$ and $a_{HHWW}$.  We find the following results for the HEFT amplitude:
 \bear
A^{\rm HEFT}&=&A_{c}^{\rm HEFT}+A_{s}^{\rm HEFT}+A_{t}^{\rm HEFT}+A_{u}^{\rm HEFT}  \nn\\
A_{c}^{\rm HEFT}&=&2  \, b \, v^{-2} \, m_W^2 \, (\epsilon_1 \cdot \epsilon_2)  \nn\\
 && +\frac{2m_W^2}{\vev^4}\left(4a_{HHWW}(s-2\mw^2)\epsilon_1\cdot\epsilon_2 -8 a_{HHWW} \epsilon_1\cdot k_2 \, \epsilon_2\cdot k_1 \right)  \nn\\
A_{s}^{\rm HEFT}&=&\frac{6 \, a \, \kappa_3 \, v^{-2} \, m_H^2 \, m_W^2 \,}{s-m_H^2 }(\epsilon_1 \cdot \epsilon_2) \nn\\
&&+ \frac{2m_W^2}{\vev^4}\frac{1}{s-\mh^2}\left( 
6\kappa_3a_{HWW}\mh^2((s-2\mw^2)\epsilon_1\cdot\epsilon_2 -2\epsilon_1\cdot k_2\,\epsilon_2\cdot k_1 ) \right) \nn \\
A_{t}^{\rm HEFT}&=&\frac{4 \, a^2 \, v^{-2} \, m_W^2}{t-m_W^2 }  \left( m_W^2 \, (\epsilon_1 \cdot \epsilon_2) + (\epsilon_1 \cdot k_3) \, (\epsilon_2 \cdot k_4) \right)  \nn\\
&&+ \frac{2m_W^2}{\vev^4}
\frac{a}{t-\mw^2} \left( -8a_{HWW}\mw^2((t+\mw^2-\mh^2)\epsilon_1\cdot\epsilon_2+\epsilon_1\cdot k_3 \, \epsilon_2\cdot k_1 +\epsilon_1\cdot k_2 \, \epsilon_2\cdot k_4) \right )  \nn\\  
A_{u}^{\rm HEFT}&=&A_{t}^{\rm HEFT} (t\leftrightarrow u,   k_3\leftrightarrow k_4) 
\label{ampWWHH-HEFT}
\eear
Regarding the important issue of gauge invariance,  it is worth commenting at this point that we have checked the gauge invariance 
of the separated contributions: $A_s$, $A_t$, $A_u$ and $A_c$.  When these contributions by channels are computed in generic $R_\xi$ gauges,  the amplitudes $A_s$ and $A_c$,  which are the ones  without vector boson propagators,  are $\xi$ independent and their values coincide in the unitary and $R_\xi$ covariant gauges.  The t-channel amplitude $A_t$  when computed in the $R_\xi$ gauges receives  contributions from two diagrams,  the  one with the $W$ propagator and the other one with the corresponding GB propagator, and when adding these two diagrams the $\xi$ dependence cancels and the result of $A_t$ is identical to the one above obtained in the unitary gauge where there is just one diagram with the $W$ propagator.  The same happens for $A_u$.  We have checked that this gauge invariance separately by channels happens in both cases,  the HEFT and the SMEFT, and for both types of scattering  with $VV=WW$ and $VV=ZZ$.  Notice also that this gauge invariance refers to all helicity amplitudes, both with initial longitudinal and transverse EW gauge bosons. 

\subsubsection{SMEFT}
The SMEFT results contain a first contribution that coincides with the SM result and a second contribution  from ${\cal L}^{(6)}$ where the corresponding coefficients appear explicitly.  We also present the SMEFT results separated by channels with the corresponding  diagrams specified in \figref{FDHH}. 
\bear
A^{\rm SMEFT}&=&A_{c}^{\rm SMEFT}+A_{s}^{\rm SMEFT}+A_{t}^{\rm SMEFT}+A_{u}^{\rm SMEFT} \nn\\
A_{c}^{\rm SMEFT}&=&2 v^{-2} \, m_W^2 \, (\epsilon_1 \cdot \epsilon_2) \nonumber\\
 &&+\quad\frac{1}{\Lambda^2} \Big(( (4 c_{\Phi \Box} - c_{\Phi D}) \, m_W^2 - 2 c_{\Phi W} \, (s-2 m_W^2) ) \, (\epsilon_1 \cdot \epsilon_2) \nonumber \\
&&+ 4 c_{\Phi W} \, (\epsilon_1 \cdot k_2) \, (\epsilon_2 \cdot k_1) \Big) \nn\\
A_{s}^{\rm SMEFT}&=& \frac{6  v^{-2} \, m_H^2 \, m_W^2 \, }{s-m_H^2} (\epsilon_1 \cdot \epsilon_2)\nonumber\\
&&-\frac{1}{(s-m_H^2)} \frac{1}{\Lambda^2} \Big( \Big( 12 \, c_{\Phi} \, m_W^2 v^2  \nonumber \\
&&- (4 c_{\Phi \Box}-c_{\Phi D}) \, m_W^2 \, (5 m_H^2 + s)  + 6 c_{\Phi W} \, m_H^2 \, (s-2 m_W^2) \Big) \, (\epsilon_1 \cdot \epsilon_2) \nonumber \\
&&- 12 \, c_{\Phi W} \, m_H^2 \, (\epsilon_1 \cdot k_2) \, (\epsilon_2 \cdot k_1) \Big) \nn\\
A_{t}^{\rm SMEFT}&=&
\frac{4 v^{-2}\, m_W^2}{t-m_W^2 }  \left( m_W^2 \, (\epsilon_1 \cdot \epsilon_2) + (\epsilon_1 \cdot k_3) \, (\epsilon_2 \cdot k_4) \right)  \nn\\
&&+\frac{2}{t-m_W^2} \frac{1}{\Lambda^2} \, m_W^2 \Big( ( (4 c_{\Phi \Box} - c_{\Phi D}) \, m_W^2 + 4 c_{\Phi W} \, (t-m_H^2 + m_W^2) ) \, (\epsilon_1 \cdot \epsilon_2) \nonumber \\
&&+ 4 c_{\Phi W} \, (\epsilon_1 \cdot k_2) ( (\epsilon_2 \cdot k_1) - (\epsilon_2 \cdot k_3) ) \nonumber \\
&&+ (\epsilon_1 \cdot k_3) ( 4 c_{\Phi W} \, (\epsilon_2 \cdot k_1) + (4 c_{\Phi \Box} - c_{\Phi D}) \, (\epsilon_2 \cdot k_4) ) \Big) \nn \\
A_{u}^{\rm SMEFT}&=&A_{t}^{\rm SMEFT} (t\leftrightarrow u,   k_3\leftrightarrow k_4)
\label{ampWWHH-SMEFT}
\eear
\subsection{$ZZ \to HH$}
We use the following momenta assignments $ Z(k_1) Z(k_2) \to H(k_3) H(k_4)$.  Now,   $\epsilon_1$ and $\epsilon_2$  are the polarization vectors of the two ingoing $Z$ bosons,  respectively.  The other conventions and simplifications used here are as in $WW \to HH$.
\subsubsection{HEFT}
As in the previous case,  for brevity,  we  keep  just the coefficients that are needed for the matching of the HEFT with the SMEFT amplitude.  We get the following results in this case:  
\bear
A^{\rm HEFT}&=&A_{c}^{\rm HEFT}+A_{s}^{\rm HEFT}+A_{t}^{\rm
HEFT}+A_{u}^{\rm HEFT} \nn \\
A_{c}^{\rm HEFT}&=&2 v^{-2}  \, b  \, m_Z^2 \, (\epsilon_1 \cdot
\epsilon_2)  \nn \\
&&
+\frac{2m_Z^2}{\vev^4}\left(4((a_{HHWW}c_W^4+a_{HHBB}s_W^4+a_{HH1}s_W^2c_W^2)(s-2\mz^2)-a_{HH0}\mz^2s_W^2)(\epsilon_1\cdot\epsilon_2)
\right.\nn\\
&&\left.-8 (a_{HHWW}c_W^4+a_{HHBB}s_W^4+a_{HH1}s_W^2c_W^2 )(\epsilon_1\cdot
k_2) \, (\epsilon_2\cdot k_1) \right)  \nn\\
A_{s}^{\rm HEFT}&=& \frac{6  v^{-2} \, a \, \kappa_3 \, m_H^2 \, m_Z^2 \,
}{s-m_H^2}(\epsilon_1 \cdot \epsilon_2)  \nn \\
&&+ \frac{2m_Z^2}{\vev^4}\frac{1}{s-\mh^2}
6\kappa_3\mh^2
\left(((a_{HWW}c_W^4+a_{HBB}s_W^4+a_{H1}s_W^2c_W^2)(s-2\mz^2)+2a_{H0}\mz^2s_W^2)(\epsilon_1\cdot\epsilon_2)
\right. \nn \\
&&\left. -2(a_{HWW}c_W^4+a_{HBB}s_W^4+a_{H1}s_W^2c_W^2)(\epsilon_1\cdot
k_2)\,(\epsilon_2\cdot k_1)  \right) \nn \\
A_{t}^{\rm HEFT}&=&\frac{4 v^{-2} \, a^2 \, m_Z^2 }{t-m_Z^2} \left( m_Z^2
\, (\epsilon_1 \cdot \epsilon_2) + (\epsilon_1 \cdot k_3) \, (\epsilon_2
\cdot k_4) \right)  \nn \\
&&+ \frac{2\mz^4}{\vev^4}
\frac{a}{t-\mz^2} \left(
-8((a_{HWW}c_W^4+a_{HBB}s_W^4+a_{H1}s_W^2c_W^2)(t+\mz^2-\mh^2)+ 2 a_{H0}\mz^2s_W^2)(\epsilon_1\cdot\epsilon_2)
\right.  \nn\\
&&\left. -8(a_{HWW}c_W^4+a_{HBB}s_W^4+a_{H1}s_W^2c_W^2)(\epsilon_1\cdot
k_3 \, \epsilon_2\cdot k_1 +\epsilon_1\cdot k_2 \, \epsilon_2\cdot k_4)
-16a_{H0} s_W^2(\epsilon_1\cdot k_3)\,(\epsilon_2\cdot k_4)\right)  \nn\\
A_{u}^{\rm HEFT}&=&A_{t}^{\rm HEFT} (t\leftrightarrow u,  
k_3\leftrightarrow k_4)
\eear
\subsubsection{SMEFT}
In the SMEFT we get the following results:
\bear
A^{\rm SMEFT}&=&A_{c}^{\rm SMEFT}+A_{s}^{\rm SMEFT}+A_{t}^{\rm SMEFT}+A_{u}^{\rm SMEFT}  \nn \\
A_{c}^{\rm SMEFT}&=&2  v^{-2}\, m_Z^2 \, (\epsilon_1 \cdot \epsilon_2) \nonumber \\
&&+ \frac{1}{\Lambda^2} \Big( 4 (c_{\Phi \Box}+c_{\Phi D}) \, m_Z^2 \, (\epsilon_1 \cdot \epsilon_2) \nonumber \\
&&\quad - 2 ( c_{\Phi B} s_W^2 +  c_{\Phi W} \, c_W^2 + c_{\Phi WB} \, s_W c_W ) \nonumber \\
&&\quad \times ( (s-2 m_Z^2 ) \, (\epsilon_1 \cdot \epsilon_2) - 2 \, (\epsilon_1 \cdot k_2) \, (\epsilon_2 \cdot k_1) ) \Big)  \nn \\
A_{s}^{\rm SMEFT}&=&\frac{6  v^{-2}\, m_H^2 \, m_Z^2 \, }{ s-m_H^2}(\epsilon_1 \cdot \epsilon_2) \nonumber\\
 &&\quad -\frac{1}{(s-m_H^2)} \frac{1}{\Lambda^2} \Big( \Big( 12 \, c_{\Phi} \, m_Z^2 v^2- 20 \, c_{\Phi \Box} \, m_H^2 \, m_Z^2 + 2 \, c_{\Phi D} \, m_H^2 \, m_Z^2 \nonumber \\
&&\quad +6m_H^2(  c_{\Phi B} \, s_W^2 +   c_{\Phi W} \, c_W^2 + c_{\Phi WB} \, s_W c_W )(s-2m_Z^2) - (4c_{\Phi \Box}-c_{\Phi D})m_Z^2 s\Big) \, (\epsilon_1 \cdot \epsilon_2) \nonumber \\
&&\quad - 12 \, m_H^2 (  c_{\Phi B} \, s_W^2 +   c_{\Phi W} \, c_W^2 + c_{\Phi WB} \, s_W c_W ) \, (\epsilon_1 \cdot k_2) \, (\epsilon_2 \cdot k_1) \Big) \nn \\
A_{t}^{\rm SMEFT}&=&\frac{4 v^{-2}\, m_Z^2}{t-m_Z^2} \left( m_Z^2 \, (\epsilon_1 \cdot \epsilon_2) + (\epsilon_1 \cdot k_3) \, (\epsilon_2 \cdot k_4) \right)\nonumber\\
&&+\frac{2 m_Z^2}{t-m_Z^2} \frac{1}{\Lambda^2} \Big( (4 c_{\Phi \Box}+c_{\Phi D} ) \, m_Z^2 \, (\epsilon_1 \cdot \epsilon_2) \nonumber \\
&&\quad + 4 \left( c_{\Phi B} s_W^2 +   c_{\Phi W} \, c_W^2 + c_{\Phi WB} \, s_W c_W  \right)\times \nonumber \\
&&\quad \times ( (t-m_H^2 + m_Z^2) \, (\epsilon_1 \cdot \epsilon_2) + (\epsilon_1 \cdot k_3) \, (\epsilon_2 \cdot k_1) \nonumber \\
&&\quad  + (\epsilon_1 \cdot k_2) \left( (\epsilon_2 \cdot k_1) - (\epsilon_2 \cdot k_3) \right) ) \nonumber \\
&&\quad + (4 c_{\Phi \Box} + c_{\Phi D})  \, (\epsilon_1 \cdot k_3) \, (\epsilon_2 \cdot k_4) \Big) \nn \\
A_{u}^{\rm SMEFT}&=&A_{t}^{\rm SMEFT} (t\leftrightarrow u,   k_3\leftrightarrow k_4) 
\eear
\newpage
\begin{figure}[!h]
    \centering
        \includegraphics[scale=0.7]{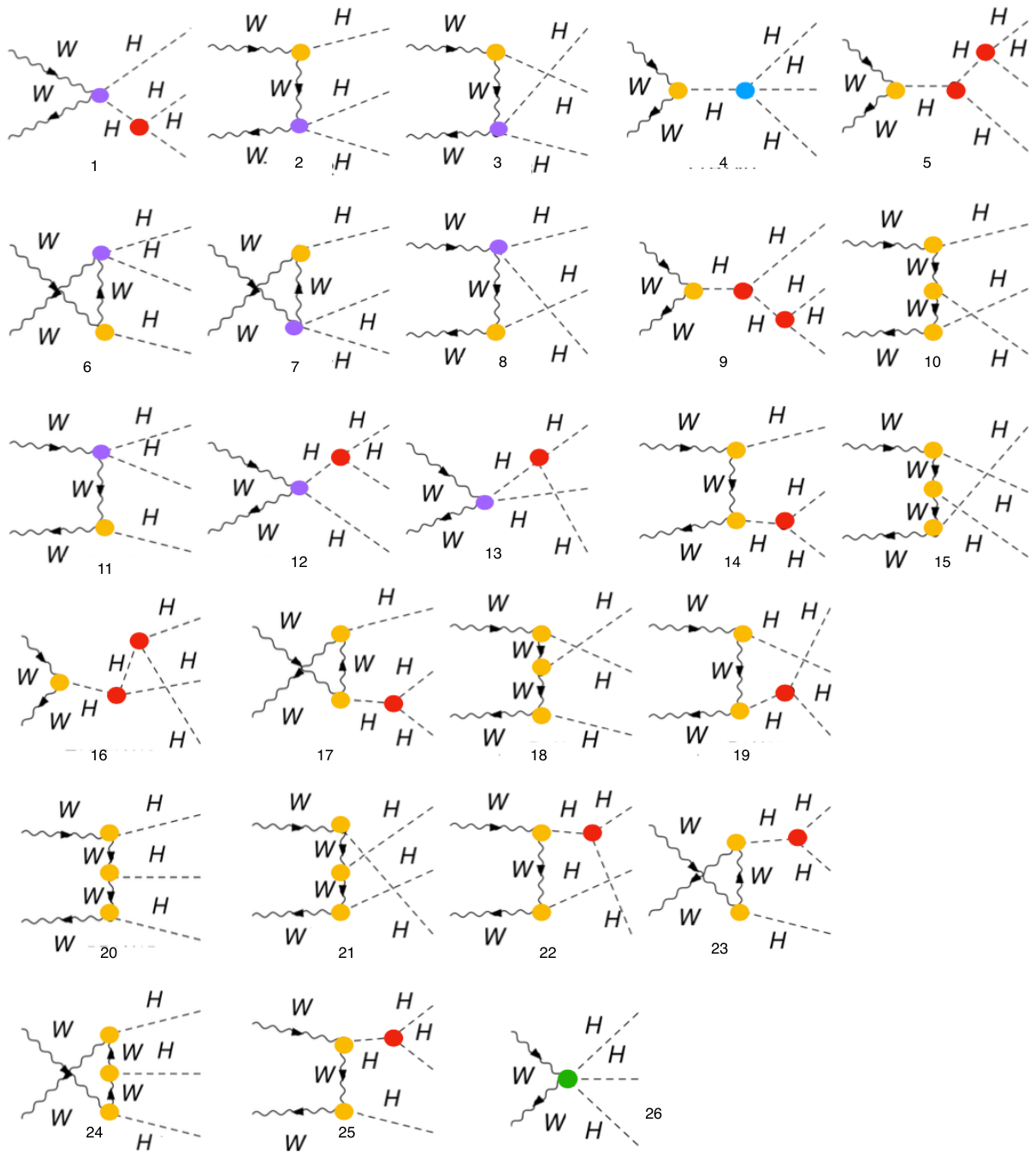}
    \caption{Feynman diagrams for $WW\to HHH$ scattering in the unitary gauge.  These diagrams are numbered from 1 to 26.  The diagrams for $ZZ \to HHH$ are similar but changing $W$ by $Z$.}
    \label{FDHHH}
    \end{figure}
\newpage

\subsection {$W W \to H H H$}
We use the following momenta assignments $ W^+(k_1)W^-(k_2) \to H(k_3) H(k_4) H(k_5)$ with $k_1$ and $k_2$ ingoing and $k_3$,  $k_4$ and $k_5$ outgoing.  The polarization vectors of the two ingoing $W$ are denoted by $\epsilon_1$ and $\epsilon_2$ respectively.  The conservation of momentum used for simplifications is then,  $k_1+k_2=k_3+k_4+k_5$. 
In this case there are 26 contributing diagrams in the unitary gauge.  They are depicted in \figref{FDHHH} with the corresponding label numbers.  In this case of triple Higgs production we  use a convenient short notation for the frequently appearing momenta combinations,  $k_{ij}=k_i-k_j$ and $q_{ij}=k_i+k_j$. 
\subsubsection{HEFT} 
Here we show the results of the HEFT, making explicit the analytical result of each diagram. 
\bear
\label{ampWWHHH-HEFT}
A^{\rm HEFT}&=&A_{1}^{\rm HEFT}+\cdots +A_{26}^{\rm HEFT} \nn \\
A_1^{\rm HEFT}&=&6 m_W^2 m_H^2 v ^{-3} \frac{ 1}{q_{45}^2-m_H^2} (b \, \kappa_3) (\epsilon_1 \cdot \epsilon_2) \nn \\
&& - 48 m_W^2 m_H^2 v^{-5}  \frac{ 1}{q_{45}^2-m_H^2} (\kappa_3 \, a_{HHWW}) ( (\epsilon_1 \cdot k_2) (\epsilon_2 \cdot k_1)- (\epsilon_1 \cdot \epsilon_2) (k_1 \cdot k_2) ) \nn \\ 
A_2^{\rm HEFT}&=&4 m_W^4  v^{-3}  \frac{ 1}{k_{13}^2-m_W^2} (a\, b)
 \left( ( \epsilon_1 \cdot \epsilon_2) -m_W^{-2} (\epsilon_1 \cdot k_{13})(\epsilon_2 \cdot k_{13}) \right) \nn \\
 && -16 m_W^4 v^{-5}  \frac{ 1}{k_{13}^2-m_W^2} \Big( 2 \, a \, a_{HHWW} ((\epsilon_1 \cdot k_2) (\epsilon_2 \cdot k_{13})- (\epsilon_1 \cdot \epsilon_2) (k_2 \cdot k_{13}) ) \nn \\
 && - \, b \, a_{HWW} ((\epsilon_2 \cdot k_1)(\epsilon_1 \cdot k_{13})-(\epsilon_1 \cdot \epsilon_2)(k_1 \cdot k_{13}) ) \Big) \nn \\
 A_3^{\rm HEFT}&=&  A_2^{\rm HEFT} (k_3 \leftrightarrow k_4) \nn\\
 A_4^{\rm HEFT}&=& 6m_W^2m_H^2v^{-3}\frac{(a \kappa_4)}{q_{12}^2-m_H^2}(\epsilon_1\cdot \epsilon_2)-24m_W^2m_H^2 v^{-5} \frac{a_{HWW} \kappa_4}{q_{12}^2-m_H^2}((\epsilon_1\cdot k_2)(\epsilon_2\cdot k_1)-(\epsilon_1\cdot \epsilon_2)(k_1\cdot k_2)) \nn\\
 A_5^{\rm HEFT}&=&  \frac{18m_W^2m_H^4 v^{-3}}{(q_{12}^2-m_H^2)(q_{34}^2-m_H^2)}\left((a \kappa_3^2)(\epsilon_1\cdot \epsilon_2)-4a_{HWW} \kappa_3^2 v^{-2}((\epsilon_1\cdot k_2)(\epsilon_2\cdot k_1)-(\epsilon_1\cdot \epsilon_2)(k_1\cdot k_2))\right)\nn\\
 A_6^{\rm HEFT}&=& A_2^{\rm HEFT} (k_3 \leftrightarrow k_5) \nn\\
 A_7^{\rm HEFT}&=& A_2^{\rm HEFT} (k_1 \leftrightarrow k_2)  \nn\\
 A_8^{\rm HEFT}&=&A_2^{\rm HEFT} (k_3 \leftrightarrow k_4,  k_1 \leftrightarrow k_2 )  \nn\\
 A_9^{\rm HEFT}&=& A_5^{\rm HEFT} (k_3 \leftrightarrow k_5)  \nn\\
 A_{10}^{\rm HEFT}&=& A_{20}^{\rm HEFT} (k_4 \leftrightarrow k_5) \nn\\
 A_{11}^{\rm HEFT}&=& A_2^{\rm HEFT} (k_3 \leftrightarrow k_5,  k_1 \leftrightarrow k_2 )  \nn\\
 A_{12}^{\rm HEFT}&=& A_1^{\rm HEFT} (k_3 \leftrightarrow k_5) \nn\\
 A_{13}^{\rm HEFT}&=& A_1^{\rm HEFT} (k_3 \leftrightarrow k_4) \nn\\
 A_{14}^{\rm HEFT}&=&12m_W^4m_H^2v^{-3}\frac{(a^2\kappa_3)}{(k_{13}^2-m_W^2)(q_{45}^2-m_H^2)}\left((\epsilon_1\cdot \epsilon_2)-m_W^{-2}(\epsilon_1\cdot k_{13})(\epsilon_2\cdot k_{13})\right)\nn\\
 &&-48{m_W^4m_H^2}v^{-5}\frac{ a\kappa_3\, a_{HWW}}{(k_{13}^2-m_W^2)(q_{45}^2-m_H^2)}\left((\epsilon_1\cdot k_2)(\epsilon_2\cdot k_{13})-(\epsilon_1\cdot \epsilon_2)(k_2\cdot k_{13})\right) \nn\\
 &&+48{m_W^4m_H^2}v^{-5}\frac{a\kappa_3\, a_{HWW}}{(k_{13}^2-m_W^2)(q_{45}^2-m_H^2)}\left((\epsilon_1\cdot k_{13})(\epsilon_2\cdot k_{1})-(\epsilon_1\cdot \epsilon_2)(k_1\cdot k_{13})\right) \nn\\
 A_{15}^{\rm HEFT}&=&  A_{20}^{\rm HEFT} ((k_3,k_4,k_5) \rightarrow (k_4,k_5,k_3)) , \nn\\
 A_{16}^{\rm HEFT}&=& A_5^{\rm HEFT} (k_4 \leftrightarrow k_5) \nn\\
 A_{17}^{\rm HEFT}&=&  A_{14}^{\rm HEFT} (k_1 \leftrightarrow k_2)  \nn\\
 A_{18}^{\rm HEFT}&=& A_{20}^{\rm HEFT} (k_3 \leftrightarrow k_4)  \nn\\
 A_{19}^{\rm HEFT}&=& A_{14}^{\rm HEFT} (k_3 \leftrightarrow k_4)  \nn\\
 A_{20}^{\rm HEFT} &=& 8 m_W^6 v^{-3}  \frac{1}{(k_{13}^2-m_W^2)(k_{25}^2-m_W^2)} (a^3) \nn \\
 && \Big( (\epsilon_1 \cdot \epsilon_2)-m_W^{-2}(\epsilon_1 \cdot k_{13}) (\epsilon_2 \cdot k_{13})-m_W^{-2}  (\epsilon_1 \cdot k_{25}) (\epsilon_2 \cdot k_{25}) \nn \\
 &&+  m_W^{-4} (\epsilon_1 \cdot k_{13}) (\epsilon_2 \cdot k_{25}) (k_{25} \cdot k_{13})  \Big) \nn \\
 && -32m_W^6 v^{-5} \frac{a^2a_{HWW}}{(k_{13}^2-m_W^2)(k_{25}^2-m_W^2)}\Big(-(\epsilon_1 \cdot k_{2})(\epsilon_2 \cdot k_{25})+(\epsilon_1 \cdot \epsilon_2)(k_2 \cdot k_{25})\nn\\&&+m_W^{-2}(\epsilon_1 \cdot k_{13})(\epsilon_2 \cdot k_{25})(k_2 \cdot k_{13})-m_W^{-2}(\epsilon_1 \cdot k_{13})(\epsilon_2 \cdot k_{13})(k_2 \cdot k_{25})\nn \\
 &&+(\epsilon_1 \cdot k_{25})(\epsilon_2 \cdot k_{13})-(\epsilon_1 \cdot \epsilon_2)(k_{13} \cdot k_{25})\nn \\
 &&-(\epsilon_1 \cdot k_{13})(\epsilon_2 \cdot k_{1})+(\epsilon_1 \cdot \epsilon_2)(k_{1} \cdot k_{13})\nn\\
&& +m_W^{-2}(\epsilon_1 \cdot k_{13})(\epsilon_2 \cdot k_{25})(k_1 \cdot k_{25})-m_W^{-2}(\epsilon_1 \cdot k_{25})(\epsilon_2 \cdot k_{25})(k_1 \cdot k_{13}) \Big)\nn \\
  A_{21}^{\rm HEFT}&=& A_{20}^{\rm HEFT} ((k_3,k_4,k_5) \rightarrow (k_5,k_3,k_4)) \nn\\
 A_{22}^{\rm HEFT}&=& A_{14}^{\rm HEFT} (k_1 \leftrightarrow k_2,  k_3 \leftrightarrow k_4) \nn\\
 A_{23}^{\rm HEFT}&=& A_{14}^{\rm HEFT} (k_3 \leftrightarrow k_5)   \nn\\
 A_{24}^{\rm HEFT}&=& A_{20}^{\rm HEFT} (k_3 \leftrightarrow k_5)\nn\\
 A_{25}^{\rm HEFT}&=& A_{14}^{\rm HEFT} (k_1 \leftrightarrow k_2,  k_3 \leftrightarrow k_5)  \nn\\
 A_{26}^{\rm HEFT}&=&6  m_W^2 v^{-3}  \,(c) \, (\epsilon_1 \cdot \epsilon_2)
\eear
Notice that we have explicitly set  some relations among various diagrams having a similar (Lorentz invariant) momenta structure  which  then,  for practical purposes,  can be grouped together into the same subset.  Within each of these subsets, two diagrams can be related to each other  by interchanges of the external momenta.  Thus,  the 26 diagrams in \figref{FDHHH}  can be grouped  into 7 subsets called here S1, S2,..., S7.   These subsets are the following;  S1: {\bf diag1},  diag12,  diag13; S2: {\bf diag2},  diag3,  diag6, diag7,  diag8, diag11; S3: {\bf diag14},  diag17,  diag19, diag22,  diag23, diag25;  S4: {\bf diag5},  diag9, diag16; S5:  diag10, diag15,  diag18,  {\bf diag20},  diag21, diag24; S6: {\bf diag4}; S7: {\bf diag26}.  In each of these subsets we have marked in boldface the diagram that has been chosen as the referent in \eqref{ampWWHHH-HEFT}.   Notice that the diagrams in each  subset can be easily visualized from \figref{FDHHH} because they all have the same pattern of colored balls in the vertices: S1 with 1 purple and 1 red; S2 with 1 purple and 1 yellow; S3 with 2 yellow and 1 red; S4 with 1 yellow and 2 red; S5 with 3 yellow; S6 with 1 yellow and 1 light blue;  S7 with 1 green.  Correspondingly,  in this $WW \to HHH$ case,  all the diagrams in the same subset contain the same combination of LO-HEFT parameters in their first terms of the above analytical results:
S1  $(b  \kappa_3)$; S2 $(a b)$;  S3  $(a^2 \kappa_3)$; S4 $(a  \kappa_3^2)$; S5  $(a^3)$;  S6 $(a  \kappa_4)$; S7 $(c)$.  As it will be shown later,  this way of grouping  the various diagrams will help in identifying the most efficient way to solve the HEFT/SMEFT matching in the forthcoming \secref{matching}.  

Finally,  regarding the gauge invariance of the above 26  amplitudes of the unitary gauge in \eqref{ampWWHHH-HEFT},  we have checked that adding together the diagrams with the same number of $W$ propagators we get identical result as in the $R_\xi$  gauges when adding to these the corresponding diagrams with the GB propagators,  and this sum leads to a $\xi$ independent result.  This is equivalent to saying that starting with the previous results for the 26 diagrams in the unitary gauge,  the sum of diagrams with 0 $W$ propagators (diag1 + diag4 + diag5 + diag9 + diag12 + diag13 + diag16 + diag26 = S1 + S4 + S6 + S7),  the sum of diagrams with 1 $W$ propagator  (diag2 + diag3 + diag6 + diag7 + diag8 + diag11 + diag14 + diag17 + diag19 + diag22 + diag23 + diag25 = S2 + S3) and the sum of diagrams with 
2 $W$ propagators (diag10 + diag15 + diag18 + diag20 + diag21 + diag24 = S5) are gauge invariant quantities separately.  

Similar classifications,  relations among diagrams and the setting of the subsets of diagrams and gauge invariant groups can be done in all the following predictions of triple Higgs production from WBF (both in HEFT and SMEFT). 

\subsubsection{SMEFT}
As in the previous case,  we write explicitly the analytical SMEFT results for each contributing diagram in \figref{FDHHH}.  Notice that we keep the same format as in the HEFT case in order to facilitate the HEFT/SMEFT comparison that will be performed when solving the matching in the forthcoming \secref{matching}. 
\bear
\label{ampWWHHH-SMEFT}
A^{\rm SMEFT}&=&A_{1}^{\rm SMEFT}+\cdots +A_{26}^{\rm SMEFT} \nn \\
A_1^{\rm SMEFT} &=& 6 m_W^2 m_H^2 v ^{-3} \frac{ 1}{q_{45}^2-m_H^2}  (\epsilon_1 \cdot \epsilon_2) \nn \\
&&-\frac{2m_W^2}{v\Lambda^2}\frac{1}{q_{45}^2-m_H^2}(6 c_\Phi v^2+(-c_{\Phi\Box}+\frac{1}{4}c_{\Phi D})(7m_H^2+2q_{45}^2))(\epsilon_1 \cdot \epsilon_2) \nn \\
&&+\frac{12m_H^2}{v\Lambda^2}\frac{1}{q_{45}^2-m_H^2}(-m_W^2(-c_{\Phi\Box}+\frac{1}{4}c_{\Phi D})(\epsilon_1\cdot \epsilon_2)+c_{\Phi W} ((\epsilon_1\cdot k_2)(\epsilon_2\cdot k_1)-(\epsilon_1\cdot \epsilon_2)( k_1\cdot k_2)))\nn \\
A_2^{\rm SMEFT}&=& 4 m_W^4  v^{-3}  \frac{ 1}{k_{13}^2-m_W^2} 
 ( ( \epsilon_1 \cdot \epsilon_2) -m_W^{-2} (\epsilon_1 \cdot k_{13})(\epsilon_2 \cdot k_{13}) ) \nn \\
 && - 12 m_W^4 \frac{1}{v\Lambda^2} \frac{ 1}{k_{13}^2-m_W^2} (-c_{\Phi\Box}+\frac{1}{4} c_{\Phi D}) ( ( \epsilon_1 \cdot \epsilon_2) -m_W^{-2} (\epsilon_1 \cdot k_{13})(\epsilon_2 \cdot k_{13}) )  \nn \\
 && + 8 m_W^2 \frac{1}{v\Lambda^2}  \frac{ 1}{k_{13}^2-m_W^2} (c_{\Phi W}) \nn \\
 &&
  \times (  (\epsilon_1 \cdot k_2) (\epsilon_2 \cdot k_{13}) -(\epsilon_1 \cdot \epsilon_2) ( k_2 \cdot k_{13}) - (\epsilon_2 \cdot k_1) (\epsilon_1 \cdot k_{13}) +
 (\epsilon_1 \cdot \epsilon_2) ( k_1 \cdot k_{13}) ) \nn \\
 A_3^{\rm SMEFT}&=&  A_2^{\rm SMEFT} (k_3 \leftrightarrow k_4) \nn\\
 A_4^{\rm SMEFT}&=&6m_W^2m_H^2 v^{-3} \frac{1}{q_{12}^2-m_H^2}(\epsilon_1\cdot \epsilon_2)\nn\\ &&-\frac{2m_W^2}{v\Lambda^2}\frac{1}{q_{12}^2-m_H^2}(36 c_\Phi v^2 +(-c_{\Phi\Box}+\frac{1}{4} c_{\Phi D})(12 m_H^2+2q_{12}^2))(\epsilon_1\cdot \epsilon_2) \nn\\
&&+\frac{6m_H^2}{v\Lambda^2}\frac{1}{q_{12}^2-m_H^2}(-m_W^2(-c_{\Phi\Box}+\frac{1}{4} c_{\Phi D})(\epsilon_1\cdot \epsilon_2)+2c_{\phi W}((\epsilon_1\cdot k_2)(\epsilon_2\cdot k_1)-(\epsilon_1\cdot \epsilon_2)(k_1\cdot k_2))) \nn\\
 A_5^{\rm SMEFT}&=& 18m_W^2m_H^4 v^{-3} \frac{1}{(q_{12}^2-m_H^2)(q_{34}^2-m_H^2)}(\epsilon_1\cdot \epsilon_2)\nn\\
&&-6m_W^2m_H^2\frac{1}{v\Lambda^2}\frac{1}{(q_{12}^2-m_H^2)(q_{34}^2-m_H^2)} \nn\\
&& \times
(12 c_\Phi v^2+(-c_{\Phi\Box}+\frac{1}{4} c_{\Phi D})(12m_H^2+2(q_{12}^2+2q_{34}^2))) (\epsilon_1\cdot \epsilon_2)\nn\\
 && +\frac{18 m_H^4}{v \Lambda^2}\frac{1}{(q_{12}^2-m_H^2)(q_{34}^2-m_H^2)} \nn\\
 && \times 
 ( -m_W^2(-c_{\Phi\Box}+ \frac{1}{4} c_{\Phi D})(\epsilon_1\cdot \epsilon_2)+2c_{\Phi W}((\epsilon_1\cdot k_2)(\epsilon_2\cdot k_1)-(\epsilon_1\cdot \epsilon_2)(k_1\cdot k_2))) \nn\\
 A_6^{\rm SMEFT}&=& A_2^{\rm SMEFT} (k_3 \leftrightarrow k_5) \nn\\
 A_7^{\rm SMEFT}&=& A_2^{\rm SMEFT} (k_1 \leftrightarrow k_2)  \nn\\
 A_8^{\rm SMEFT}&=&A_2^{\rm SMEFT} (k_3 \leftrightarrow k_4,  k_1 \leftrightarrow k_2 )  \nn\\
 A_9^{\rm SMEFT}&=& A_5^{\rm SMEFT} (k_3 \leftrightarrow k_5)  \nn\\
 A_{10}^{\rm SMEFT}&=& A_{20}^{\rm SMEFT} (k_4 \leftrightarrow k_5) \nn\\
 A_{11}^{\rm SMEFT}&=& A_2^{\rm SMEFT} (k_3 \leftrightarrow k_5,  k_1 \leftrightarrow k_2 )  \nn\\
 A_{12}^{\rm SMEFT}&=& A_1^{\rm SMEFT} (k_3 \leftrightarrow k_5) \nn\\
 A_{13}^{\rm SMEFT}&=& A_1^{\rm SMEFT} (k_3 \leftrightarrow k_4) \nn\\
 A_{14}^{\rm SMEFT}&=&12 m_W^4 m_H^2v^{-3}\frac{1}{(k_{13}^2-m_W^2)(q_{45}^2-m_H^2)}\left((\epsilon_1\cdot \epsilon_2)-m_W^{-2}(\epsilon_1\cdot k_{13})(\epsilon_2\cdot k_{13}) \right) \nn\\
  &&-\frac{4m_W^4}{v\Lambda^2}\frac{1}{(k_{13}^2-m_W^2)(q_{45}^2-m_H^2)} \nn \\
  && \times
  (6c_{\Phi} v^2+(-c_{\Phi\Box}+\frac{1}{4} c_{\Phi D})(7m_H^2+2q_{45}^2))\left((\epsilon_1\cdot \epsilon_2)-m_W^{-2}(\epsilon_1\cdot k_{13})(\epsilon_2\cdot k_{13}) \right) \nn\\
  && +\frac{12m_W^2m_H^2}{v\Lambda^2}\frac{1}{(k_{13}^2-m_W^2)(q_{45}^2-m_H^2)}\Big(-2m_W^2( -c_{\Phi\Box}+\frac{1}{4} c_{\Phi D})((\epsilon_1\cdot \epsilon_2)-m_W^{-2}(\epsilon_1\cdot k_{13})(\epsilon_2\cdot k_{13}) )\nn\\
  &&+2c_{\phi W}((\epsilon_1\cdot k_{2})(\epsilon_2\cdot k_{13})-(\epsilon_1\cdot \epsilon_2)(k_2\cdot k_{13})-(\epsilon_1\cdot k_{13})(\epsilon_2\cdot k_{1})+(\epsilon_1\cdot \epsilon_2)(k_1\cdot k_{13})))\Big)\nn\\
   A_{15}^{\rm SMEFT}&=&  A_{20}^{\rm SMEFT} ((k_3,k_4,k_5) \rightarrow (k_4,k_5,k_3))\nn\\
 A_{16}^{\rm SMEFT}&=& A_5^{\rm SMEFT} (k_4 \leftrightarrow k_5) \nn\\
 A_{17}^{\rm SMEFT}&=&  A_{14}^{\rm SMEFT} (k_1 \leftrightarrow k_2)  \nn\\
 A_{18}^{\rm SMEFT}&=& A_{20}^{\rm SMEFT} (k_3 \leftrightarrow k_4)  \nn\\
 A_{19}^{\rm SMEFT}&=& A_{14}^{\rm SMEFT} (k_3 \leftrightarrow k_4)  \nn\\
 A_{20}^{\rm SMEFT} &=& 8 m_W^6 v^{-3}  \frac{1}{(k_{13}^2-m_W^2)(k_{25}^2-m_W^2)} \Big( (\epsilon_1 \cdot \epsilon_2)-m_W^{-2}(\epsilon_1 \cdot k_{13}) (\epsilon_2 \cdot k_{13})-m_W^{-2}  (\epsilon_1 \cdot k_{25}) (\epsilon_2 \cdot k_{25}) \nn \\
 &&+  m_W^{-4} (\epsilon_1 \cdot k_{13}) (\epsilon_2 \cdot k_{25}) (k_{25} \cdot k_{13})  \Big) \nn \\
 && + \frac{8m_W^4}{v\Lambda^2}\frac{1}{(k_{13}^2-m_W^2)(k_{25}^2-m_W^2)}\Big(-3m^2_W(-c_{\Phi\Box}+\frac{1}{4} c_{\Phi D})\big((\epsilon_1 \cdot \epsilon_2)-m_W^{-2}(\epsilon_1 \cdot k_{25})(\epsilon_2 \cdot k_{25})\nn\\
 && -m_W^{-2}(\epsilon_1 \cdot k_{13})(\epsilon_2 \cdot k_{13})-m_W^{-4}(\epsilon_1 \cdot k_{13})(\epsilon_2 \cdot k_{25})(k_{13} \cdot k_{25})\big)\nn \\
&& + 2c_{\Phi W}\big( -(\epsilon_1 \cdot k_{2})(\epsilon_2 \cdot k_{25})+(\epsilon_1 \cdot \epsilon_{2})(k_2 \cdot k_{25})\nn \\
&&+m_W^{-2}(\epsilon_1 \cdot k_{13})(\epsilon_2 \cdot k_{25})(k_2 \cdot k_{13})-m^{-2}_W (\epsilon_1 \cdot k_{13}) (\epsilon_2 \cdot k_{13})(k_2 \cdot k_{25}) \nn\\
&&+(\epsilon_1 \cdot k_{25})(\epsilon_2 \cdot k_{13})-(\epsilon_1 \cdot \epsilon_{2})(k_{13} \cdot k_{25})-(\epsilon_1 \cdot k_{13})(\epsilon_2 \cdot k_{1})+(\epsilon_1 \cdot \epsilon_{2})(k_1 \cdot k_{13})\nn \\
&&+m_W^{-2}(\epsilon_1 \cdot k_{13})(\epsilon_2 \cdot k_{25})(k_1 \cdot k_{25})-m^{-2}_W (\epsilon_1 \cdot k_{25}) (\epsilon_2 \cdot k_{25})(k_1 \cdot k_{13})\big)\Big) \nn \\
 A_{21}^{\rm SMEFT}&=& A_{20}^{\rm SMEFT} ((k_3,k_4,k_5) \rightarrow (k_5,k_3,k_4)) \nn\\
 A_{22}^{\rm SMEFT}&=& A_{14}^{\rm SMEFT} (k_1 \leftrightarrow k_2,  k_3 \leftrightarrow k_4) \nn\\
 A_{23}^{\rm SMEFT}&=& A_{14}^{\rm SMEFT} (k_3 \leftrightarrow k_5)   \nn\\
 A_{24}^{\rm SMEFT}&=& A_{20}^{\rm SMEFT} (k_3 \leftrightarrow k_5)\nn\\
 A_{25}^{\rm SMEFT}&=& A_{14}^{\rm SMEFT} (k_1 \leftrightarrow k_2,  k_3 \leftrightarrow k_5)  \nn\\ 
A_{26}^{\rm SMEFT} &=&  0
\eear

\subsection {$ZZ \to H H H$}
We use the following momenta assignments $ Z(k_1)Z(k_2) \to H(k_3) H(k_4) H(k_5)$.  The polarization vectors of the two ingoing $Z$ are denoted by $\epsilon_1$ and $\epsilon_2$ respectively. The other conventions and simplifications used here are as in $WW \to HHH$.  In this case of triple Higgs production we  use  again the short notation for the frequently appearing momenta combinations,  $k_{ij}=k_i-k_j$ and $q_{ij}=k_i+k_j$.

\subsubsection{HEFT}
Here we show the results of the HEFT, making explicit the analytical result of each diagram. 
\bear
\label{ampZZHHH-HEFT}
A^{\rm HEFT}&=&A_{1}^{\rm HEFT}+\cdots +A_{26}^{\rm HEFT} \nn \\
A_1^{\rm HEFT}&=&6 m_Z^2 m_H^2 v ^{-3} \frac{ 1}{q_{45}^2-m_H^2} (b \, \kappa_3) (\epsilon_1 \cdot \epsilon_2) \nn \\
&& - 48 m_Z^2 m_H^2 v^{-5}  \frac{ 1}{q_{45}^2-m_H^2} (\kappa_3)( a_{HHWW}c_W^4+a_{HHBB} s_W^4+a_{HH1} s_W^2 c_W^2) \nn \\ 
&&\times ( (\epsilon_1 \cdot k_2) (\epsilon_2 \cdot k_1)- (\epsilon_1 \cdot \epsilon_2) (k_1 \cdot k_2) )- 48 m_Z^4 m_H^2 v^{-5}  \frac{ 1}{q_{45}^2-m_H^2} (\kappa_3 a_{HH0} s_W^2)(\epsilon_1 \cdot \epsilon_2)\nn\\
A_2^{\rm HEFT}&=&4 m_Z^4  v^{-3}  \frac{ 1}{k_{13}^2-m_Z^2} (a\, b)
 \left( ( \epsilon_1 \cdot \epsilon_2) -m_Z^{-2} (\epsilon_1 \cdot k_{13})(\epsilon_2 \cdot k_{13}) \right) \nn \\
 && -16 m_Z^4 v^{-5}  \frac{ 1}{k_{13}^2-m_Z^2} \Big( 2 \, a \, ( a_{HHWW}c_W^4+a_{HHBB} s_W^4+a_{HH1} s_W^2 c_W^2)\nn\\
 && \times ((\epsilon_1 \cdot k_2) (\epsilon_2 \cdot k_{13})- (\epsilon_1 \cdot \epsilon_2) (k_2 \cdot k_{13}) ) \nn \\
 && - \, b \, ( a_{HWW}c_W^4+a_{HBB} s_W^4+a_{H1} s_W^2 c_W^2) ((\epsilon_2 \cdot k_1)(\epsilon_1 \cdot k_{13})-(\epsilon_1 \cdot \epsilon_2)(k_1 \cdot k_{13}) ) \Big) \nn \\
 && -16 m_Z^6 v^{-5}  \frac{ 1}{k_{13}^2-m_Z^2}  (2 \, a \, a_{HH0}s^2_W - \, b \, a_{H0}s^2_W )  \left( ( \epsilon_1 \cdot \epsilon_2) -m_Z^{-2} (\epsilon_1 \cdot k_{13})(\epsilon_2 \cdot k_{13}) \right)\nn \\
 A_3^{\rm HEFT}&=&  A_2^{\rm HEFT} (k_3 \leftrightarrow k_4) \nn\\
 A_4^{\rm HEFT}&=& 6m_Z^2m_H^2v^{-3}\frac{a \kappa_4}{q_{12}^2-m_H^2}(\epsilon_1\cdot \epsilon_2)-24m_Z^2m_H^2 v^{-5} \frac{1}{q_{12}^2-m_H^2}\nn\\
 &&\times \kappa_4( a_{HWW}c_W^4+a_{HBB} s_W^4+a_{H1} s_W^2 c_W^2)((\epsilon_1\cdot k_2)(\epsilon_2\cdot k_1)-(\epsilon_1\cdot \epsilon_2)(k_1\cdot k_2)) \nn\\
 &&-24m_Z^4m_H^2 v^{-5} \frac{\kappa_4a_{H0}s_W^2}{q_{12}^2-m_H^2}(\epsilon_1\cdot \epsilon_2)\nn\\
 A_5^{\rm HEFT}&=&  \frac{18m_Z^2m_H^4 v^{-3}}{(q_{12}^2-m_H^2)(q_{34}^2-m_H^2)}\Big(a \kappa_3^2(\epsilon_1\cdot \epsilon_2)-4\kappa_3^2 v^{-2}( a_{HWW}c_W^4+a_{HBB} s_W^4+a_{H1} s_W^2 c_W^2) \nn\\
 &&\times((\epsilon_1\cdot k_2)(\epsilon_2\cdot k_1)-(\epsilon_1\cdot \epsilon_2)(k_1\cdot k_2))-4m_Z^2v^{-2}\kappa_3^2a_{H0}s_W^2(\epsilon_1\cdot \epsilon_2)\Big)\nn\\
 A_6^{\rm HEFT}&=& A_2^{\rm HEFT} (k_3 \leftrightarrow k_5) \nn\\
 A_7^{\rm HEFT}&=& A_2^{\rm HEFT} (k_1 \leftrightarrow k_2)  \nn\\
 A_8^{\rm HEFT}&=&A_2^{\rm HEFT} (k_3 \leftrightarrow k_4,  k_1 \leftrightarrow k_2 )  \nn\\
 A_9^{\rm HEFT}&=& A_5^{\rm HEFT} (k_3 \leftrightarrow k_5)  \nn\\
 A_{10}^{\rm HEFT}&=& A_{20}^{\rm HEFT} (k_4 \leftrightarrow k_5) \nn\\
 A_{11}^{\rm HEFT}&=& A_2^{\rm HEFT} (k_3 \leftrightarrow k_5,  k_1 \leftrightarrow k_2 )  \nn\\
 A_{12}^{\rm HEFT}&=& A_1^{\rm HEFT} (k_3 \leftrightarrow k_5) \nn\\
 A_{13}^{\rm HEFT}&=& A_1^{\rm HEFT} (k_3 \leftrightarrow k_4) \nn\\
 A_{14}^{\rm HEFT}&=&12m_Z^4m_H^2v^{-3}\frac{a^2\kappa_3}{(k_{13}^2-m_Z^2)(q_{45}^2-m_H^2)}\left((\epsilon_1\cdot \epsilon_2)-m_Z^{-2}(\epsilon_1\cdot k_{13})(\epsilon_2\cdot k_{13})\right)\nn\\
 &&-48{m_Z^4m_H^2}v^{-5}\frac{a\kappa_3( a_{HWW}c_W^4+a_{HBB} s_W^4+a_{H1} s_W^2 c_W^2) }{(k_{13}^2-m_Z^2)(q_{45}^2-m_H^2)}\left((\epsilon_1\cdot k_2)(\epsilon_2\cdot k_{13})-(\epsilon_1\cdot \epsilon_2)(k_2\cdot k_{13})\right) \nn\\
 &&+48{m_Z^4m_H^2}v^{-5}\frac{a\kappa_3( a_{HWW}c_W^4+a_{HBB} s_W^4+a_{H1} s_W^2 c_W^2) }{(k_{13}^2-m_Z^2)(q_{45}^2-m_H^2)}\left((\epsilon_1\cdot k_{13})(\epsilon_2\cdot k_{1})-(\epsilon_1\cdot \epsilon_2)(k_1\cdot k_{13})\right) \nn\\
 &&-96{m_Z^6m_H^2}v^{-5}\frac{a\kappa_3\,a_{H0}s_W^2 }{(k_{13}^2-m_Z^2)(q_{45}^2-m_H^2)}\left((\epsilon_1\cdot \epsilon_2)-m_Z^{-2}(\epsilon_1\cdot k_{13})(\epsilon_2\cdot k_{13})\right) \nn\\
 A_{15}^{\rm HEFT}&=&  A_{20}^{\rm HEFT} ((k_3,k_4,k_5) \rightarrow (k_4,k_5,k_3)) , \nn\\
 A_{16}^{\rm HEFT}&=& A_5^{\rm HEFT} (k_4 \leftrightarrow k_5) \nn\\
 A_{17}^{\rm HEFT}&=&  A_{14}^{\rm HEFT} (k_1 \leftrightarrow k_2)  \nn\\
 A_{18}^{\rm HEFT}&=& A_{20}^{\rm HEFT} (k_3 \leftrightarrow k_4)  \nn\\
 A_{19}^{\rm HEFT}&=& A_{14}^{\rm HEFT} (k_3 \leftrightarrow k_4)  \nn\\
 A_{20}^{\rm HEFT} &=& 8 m_Z^6 v^{-3}  \frac{1}{(k_{13}^2-m_Z^2)(k_{25}^2-m_Z^2)} (a^3) \nn \\
 && \times\Big( (\epsilon_1 \cdot \epsilon_2)-m_Z^{-2}(\epsilon_1 \cdot k_{13}) (\epsilon_2 \cdot k_{13})-m_Z^{-2}  (\epsilon_1 \cdot k_{25}) (\epsilon_2 \cdot k_{25}) \nn \\
 &&+  m_Z^{-4} (\epsilon_1 \cdot k_{13}) (\epsilon_2 \cdot k_{25}) (k_{25} \cdot k_{13})  \Big) \nn \\
 && -32m_Z^6 v^{-5} \frac{a^2( a_{HWW}c_W^4+a_{HBB} s_W^4+a_{H1} s_W^2 c_W^2)}{(k_{13}^2-m_Z^2)(k_{25}^2-m_Z^2)}\Big(-(\epsilon_1 \cdot k_{2})(\epsilon_2 \cdot k_{25})\nn\\&&+(\epsilon_1 \cdot \epsilon_2)(k_2 \cdot k_{25})+m_Z^{-2}(\epsilon_1 \cdot k_{13})(\epsilon_2 \cdot k_{25})(k_2 \cdot k_{13})-m_Z^{-2}(\epsilon_1 \cdot k_{13})(\epsilon_2 \cdot k_{13})(k_2 \cdot k_{25})\nn \\
 &&+(\epsilon_1 \cdot k_{25})(\epsilon_2 \cdot k_{13})-(\epsilon_1 \cdot \epsilon_2)(k_{13} \cdot k_{25})\nn \\
 &&-(\epsilon_1 \cdot k_{13})(\epsilon_2 \cdot k_{1})+(\epsilon_1 \cdot \epsilon_2)(k_{1} \cdot k_{13})\nn\\
&& +m_Z^{-2}(\epsilon_1 \cdot k_{13})(\epsilon_2 \cdot k_{25})(k_1 \cdot k_{25})-m_W^{-2}(\epsilon_1 \cdot k_{25})(\epsilon_2 \cdot k_{25})(k_1 \cdot k_{13}) \Big)\nn \\
&&-96m_Z^8 v^{-5}\frac{a^2 a_{H0}s_W^2}{(k_{13}^2-m_Z^2)(k_{25}^2-m_Z^2)} \Big( (\epsilon_1 \cdot \epsilon_2)-m_Z^{-2}(\epsilon_1 \cdot k_{13}) (\epsilon_2 \cdot k_{13}) \nn \\
 &&-m_Z^{-2}  (\epsilon_1 \cdot k_{25}) (\epsilon_2 \cdot k_{25})+  m_Z^{-4} (\epsilon_1 \cdot k_{13}) (\epsilon_2 \cdot k_{25}) (k_{25} \cdot k_{13})  \Big)\nn\\
  A_{21}^{\rm HEFT}&=& A_{20}^{\rm HEFT} ((k_3,k_4,k_5) \rightarrow (k_5,k_3,k_4)) \nn\\
 A_{22}^{\rm HEFT}&=& A_{14}^{\rm HEFT} (k_1 \leftrightarrow k_2,  k_3 \leftrightarrow k_4) \nn\\
 A_{23}^{\rm HEFT}&=& A_{14}^{\rm HEFT} (k_3 \leftrightarrow k_5)   \nn\\
 A_{24}^{\rm HEFT}&=& A_{20}^{\rm HEFT} (k_3 \leftrightarrow k_5)\nn\\
 A_{25}^{\rm HEFT}&=& A_{14}^{\rm HEFT} (k_1 \leftrightarrow k_2,  k_3 \leftrightarrow k_5)  \nn\\
 A_{26}^{\rm HEFT}&=&6  m_W^2 v^{-3}  \,(c) \, (\epsilon_1 \cdot \epsilon_2)  -48m_Z^4 v^{-5} a_{HHH0}s_W^2\, (\epsilon_1 \cdot \epsilon_2)
\eear

\subsubsection{SMEFT}
As in the previous case,  we write explicitly the analytical SMEFT results for each contributing diagram in \figref{FDHHH}.
\bear
\label{ampZZHHH-SMEFT}
A^{\rm SMEFT}&=&A_{1}^{\rm SMEFT}+\cdots +A_{26}^{\rm SMEFT} \nn \\
A_1^{\rm SMEFT} &=& 6 m_Z^2 m_H^2 v ^{-3} \frac{ 1}{q_{45}^2-m_H^2}  (\epsilon_1 \cdot \epsilon_2) \nn \\
&&-\frac{2m_Z^2}{v\Lambda^2}\frac{1}{q_{45}^2-m_H^2}(6 c_\Phi v^2+(-c_{\Phi\Box}+\frac{1}{4}c_{\Phi D})(7m_H^2+2q_{45}^2))(\epsilon_1 \cdot \epsilon_2) \nn \\
&&+\frac{12m_Z^2}{v\Lambda^2}\frac{1}{q_{45}^2-m_H^2}\Big(-m_Z^2(-c_{\Phi\Box}+\frac{1}{4}c_{\Phi D})(\epsilon_1\cdot \epsilon_2)+(c_{\Phi W} c_W^2+c_{\Phi B} s_W^2+c_{\Phi WB} c_Ws_W)\nn \\
&&\times ((\epsilon_1\cdot k_2)(\epsilon_2\cdot k_1)-(\epsilon_1\cdot \epsilon_2)( k_1\cdot k_2))\Big)\nn \\
&&+\frac{15 m_Z^2 m_H^2}{v\Lambda^2} \frac{ 1}{q_{45}^2-m_H^2} c_{\Phi D}  (\epsilon_1 \cdot \epsilon_2)\nn\\
A_2^{\rm SMEFT}&=& 4 m_Z^4  v^{-3}  \frac{ 1}{k_{13}^2-m_Z^2} 
 ( ( \epsilon_1 \cdot \epsilon_2) -m_Z^{-2} (\epsilon_1 \cdot k_{13})(\epsilon_2 \cdot k_{13}) ) \nn \\
 && - 12 m_Z^4 \frac{1}{v\Lambda^2} \frac{ 1}{k_{13}^2-m_Z^2} (-c_{\Phi\Box}+\frac{1}{4} c_{\Phi D}) ( ( \epsilon_1 \cdot \epsilon_2) -m_Z^{-2} (\epsilon_1 \cdot k_{13})(\epsilon_2 \cdot k_{13}) )  \nn \\
 && + 8 m_Z^2 \frac{1}{v\Lambda^2}  \frac{ 1}{k_{13}^2-m_Z^2} (c_{\Phi W}c_W^2+c_{\Phi B}s_W^2+c_{\Phi WB}c_Ws_W) \nn \\
 &&
  \times (  (\epsilon_1 \cdot k_2) (\epsilon_2 \cdot k_{13}) -(\epsilon_1 \cdot \epsilon_2) ( k_2 \cdot k_{13}) - (\epsilon_2 \cdot k_1) (\epsilon_1 \cdot k_{13}) +
 (\epsilon_1 \cdot \epsilon_2) ( k_1 \cdot k_{13}) ) \nn \\
 &&+\frac{12m_Z^4}{v \Lambda^2} \frac{ 1}{k_{13}^2-m_Z^2} (c_{\Phi D})( ( \epsilon_1 \cdot \epsilon_2) -m_Z^{-2} (\epsilon_1 \cdot k_{13})(\epsilon_2 \cdot k_{13}) )\nn\\
 A_3^{\rm SMEFT}&=&  A_2^{\rm SMEFT} (k_3 \leftrightarrow k_4) \nn\\
 A_4^{\rm SMEFT}&=&6m_Z^2m_H^2 v^{-3} \frac{1}{q_{12}^2-m_H^2}(\epsilon_1\cdot \epsilon_2)\nn\\ 
 &&-\frac{2m_Z^2}{v\Lambda^2}\frac{1}{q_{12}^2-m_H^2}(36 c_\Phi v^2 +(-c_{\Phi\Box}+\frac{1}{4} c_{\Phi D})(12 m_H^2+2q_{12}^2))(\epsilon_1\cdot \epsilon_2) \nn\\
&&+\frac{6m_H^2}{v\Lambda^2}\frac{1}{q_{12}^2-m_H^2}\Big(-m_Z^2(-c_{\Phi\Box}+\frac{1}{4} c_{\Phi D})(\epsilon_1\cdot \epsilon_2)+2(c_{\Phi W}c_W^2+c_{\Phi B}s_W^2+c_{\Phi WB}c_Ws_W)\nn\\
&&\times((\epsilon_1\cdot k_2)(\epsilon_2\cdot k_1)-(\epsilon_1\cdot \epsilon_2)(k_1\cdot k_2))\Big)+\frac{3m_Z^2m_H^2}{v\Lambda^2}\frac{1}{q_{12}^2-m_H^2}c_{\Phi D} (\epsilon_1\cdot \epsilon_2) \nn\\
 A_5^{\rm SMEFT}&=& 18m_Z^2m_H^4 v^{-3} \frac{1}{(q_{12}^2-m_H^2)(q_{34}^2-m_H^2)}(\epsilon_1\cdot \epsilon_2)\nn\\
&&-6m_Z^2m_H^2\frac{1}{v\Lambda^2}\frac{1}{(q_{12}^2-m_H^2)(q_{34}^2-m_H^2)} \nn \\
&& \times
(12 c_\Phi v^2+(-c_{\Phi\Box}+\frac{1}{4} c_{\Phi D})(12m_H^2+2(q_{12}^2+2q_{34}^2))) (\epsilon_1\cdot \epsilon_2)\nn\\
 && +\frac{18 m_H^4}{v \Lambda^2}\frac{1}{(q_{12}^2-m_H^2)(q_{34}^2-m_H^2)}\Big( -m_Z^2( -c_{\Phi\Box}+\frac{1}{4} c_{\Phi D})(\epsilon_1\cdot \epsilon_2)\nn\\
 &&+2(c_{\Phi W}c_W^2+c_{\Phi B}s_W^2+c_{\Phi WB}c_Ws_W)((\epsilon_1\cdot k_2)(\epsilon_2\cdot k_1)-(\epsilon_1\cdot \epsilon_2)(k_1\cdot k_2))\Big) \nn\\
 && +\frac{9m_Z^2 m_H^2}{v\Lambda^2}\frac{1}{(q_{12}^2-m_H^2)(q_{34}^2-m_H^2)}c_{\Phi D}(\epsilon_1\cdot \epsilon_2) \nn\\
 A_6^{\rm SMEFT}&=& A_2^{\rm SMEFT} (k_3 \leftrightarrow k_5) \nn\\
 A_7^{\rm SMEFT}&=& A_2^{\rm SMEFT} (k_1 \leftrightarrow k_2)  \nn\\
 A_8^{\rm SMEFT}&=&A_2^{\rm SMEFT} (k_3 \leftrightarrow k_4,  k_1 \leftrightarrow k_2 )  \nn\\
 A_9^{\rm SMEFT}&=& A_5^{\rm SMEFT} (k_3 \leftrightarrow k_5)  \nn\\
 A_{10}^{\rm SMEFT}&=& A_{20}^{\rm SMEFT} (k_4 \leftrightarrow k_5) \nn\\
 A_{11}^{\rm SMEFT}&=& A_2^{\rm SMEFT} (k_3 \leftrightarrow k_5,  k_1 \leftrightarrow k_2 )  \nn\\
 A_{12}^{\rm SMEFT}&=& A_1^{\rm SMEFT} (k_3 \leftrightarrow k_5) \nn\\
 A_{13}^{\rm SMEFT}&=& A_1^{\rm SMEFT} (k_3 \leftrightarrow k_4) \nn\\
 A_{14}^{\rm SMEFT}&=&12 m_Z^4 m_H^2v^{-3}\frac{1}{(k_{13}^2-m_Z^2)(q_{45}^2-m_H^2)}\left((\epsilon_1\cdot \epsilon_2)-m_Z^{-2}(\epsilon_1\cdot k_{13})(\epsilon_2\cdot k_{13}) \right) \nn\\
  &&-\frac{4m_Z^4}{v\Lambda^2}\frac{1}{(k_{13}^2-m_Z^2)(q_{45}^2-m_H^2)} \nn \\
  && \times
  (6c_{\Phi} v^2+(-c_{\Phi\Box}+\frac{1}{4} c_{\Phi D})(7m_H^2+2q_{45}^2))\left((\epsilon_1\cdot \epsilon_2)-m_Z^{-2}(\epsilon_1\cdot k_{13})(\epsilon_2\cdot k_{13}) \right) \nn\\
  && +\frac{12m_Z^2m_H^2}{v\Lambda^2}\frac{1}{(k_{13}^2-m_Z^2)(q_{45}^2-m_H^2)}\Big(-2m_Z^2( -c_{\Phi\Box}+\frac{1}{4} c_{\Phi D})((\epsilon_1\cdot \epsilon_2)-m_W^{-2}(\epsilon_1\cdot k_{13})(\epsilon_2\cdot k_{13}) )\nn\\
  &&+2(c_{\Phi W}c_W^2+c_{\Phi B}s_W^2+c_{\Phi WB}c_Ws_W)\nn\\
  &&\times((\epsilon_1\cdot k_{2})(\epsilon_2\cdot k_{13})-(\epsilon_1\cdot \epsilon_2)(k_2\cdot k_{13})-(\epsilon_1\cdot k_{13})(\epsilon_2\cdot k_{1})+(\epsilon_1\cdot \epsilon_2)(k_1\cdot k_{13})\Big)\nn\\
  &&+\frac{12m_Z^4 m_H^2}{v \Lambda^2}\frac{1}{(k_{13}^2-m_Z^2)(q_{45}^2-m_H^2)} c_{\Phi D} \left((\epsilon_1\cdot \epsilon_2)-m_Z^{-2}(\epsilon_1\cdot k_{13})(\epsilon_2\cdot k_{13}) \right) \nn\\
   A_{15}^{\rm SMEFT}&=&  A_{20}^{\rm SMEFT} ((k_3,k_4,k_5) \rightarrow (k_4,k_5,k_3))\nn\\
 A_{16}^{\rm SMEFT}&=& A_5^{\rm SMEFT} (k_4 \leftrightarrow k_5) \nn\\
 A_{17}^{\rm SMEFT}&=&  A_{14}^{\rm SMEFT} (k_1 \leftrightarrow k_2)  \nn\\
 A_{18}^{\rm SMEFT}&=& A_{20}^{\rm SMEFT} (k_3 \leftrightarrow k_4)  \nn\\
 A_{19}^{\rm SMEFT}&=& A_{14}^{\rm SMEFT} (k_3 \leftrightarrow k_4)  \nn\\
 A_{20}^{\rm SMEFT} &=& 8 m_Z^6 v^{-3}  \frac{1}{(k_{13}^2-m_Z^2)(k_{25}^2-m_Z^2)} \Big( (\epsilon_1 \cdot \epsilon_2)-m_Z^{-2}(\epsilon_1 \cdot k_{13}) (\epsilon_2 \cdot k_{13})-m_Z^{-2}  (\epsilon_1 \cdot k_{25}) (\epsilon_2 \cdot k_{25}) \nn \\
 &&+  m_Z^{-4} (\epsilon_1 \cdot k_{13}) (\epsilon_2 \cdot k_{25}) (k_{25} \cdot k_{13})  \Big) \nn \\
 && + \frac{8m_Z^4}{v\Lambda^2}\frac{1}{(k_{13}^2-m_Z^2)(k_{25}^2-m_Z^2)}\Big(-3(-c_{\Phi\Box}+\frac{1}{4} c_{\Phi D})\big(m^2_Z(\epsilon_1 \cdot \epsilon_2)-(\epsilon_1 \cdot k_{25})(\epsilon_2 \cdot k_{25})\nn\\
 && -(\epsilon_1 \cdot k_{13})(\epsilon_2 \cdot k_{13})+m_Z^{-2}(\epsilon_1 \cdot k_{13})(\epsilon_2 \cdot k_{25})(k_{13} \cdot k_{25})\big)\nn \\
&& + 2(c_{\Phi W}c_W^2+c_{\Phi B}s_W^2+c_{\Phi WB}c_Ws_W)\big((\epsilon_1 \cdot k_{25})(\epsilon_2 \cdot k_{13})-(\epsilon_1 \cdot \epsilon_{2})(k_{13} \cdot k_{25})  \nn\\
&&-(\epsilon_1 \cdot k_{13})(\epsilon_2 \cdot k_{1})+(\epsilon_1 \cdot \epsilon_{2})(k_1 \cdot k_{13})+m_Z^{-2}(\epsilon_1 \cdot k_{13})(\epsilon_2 \cdot k_{25})(k_1 \cdot k_{25})\nn \\
&&-m^{-2}_Z (\epsilon_1 \cdot k_{25}) (\epsilon_2 \cdot k_{25})(k_1 \cdot k_{13})\big) -(\epsilon_1 \cdot k_{2})(\epsilon_2 \cdot k_{25})+(\epsilon_1 \cdot \epsilon_{2})(k_2 \cdot k_{25})\nn\\
&&+m_Z^{-2}(\epsilon_1 \cdot k_{13})(\epsilon_2 \cdot k_{25})(k_2 \cdot k_{13})-m^{-2}_Z (\epsilon_1 \cdot k_{13}) (\epsilon_2 \cdot k_{13})(k_2 \cdot k_{25})\big)\Big) \nn \\
&&+\frac{12m_Z^6}{v \Lambda^2}\frac{1}{(k_{13}^2-m_Z^2)(k_{25}^2-m_Z^2)} c_{\Phi D}\Big( (\epsilon_1 \cdot \epsilon_2)-m_Z^{-2}(\epsilon_1 \cdot k_{13}) (\epsilon_2 \cdot k_{13}) \nn \\
 &&-m_Z^{-2}  (\epsilon_1 \cdot k_{25}) (\epsilon_2 \cdot k_{25})+  m_Z^{-4} (\epsilon_1 \cdot k_{13}) (\epsilon_2 \cdot k_{25}) (k_{25} \cdot k_{13})  \Big) \nn\\
 A_{21}^{\rm SMEFT}&=& A_{20}^{\rm SMEFT} ((k_3,k_4,k_5) \rightarrow (k_5,k_3,k_4)) \nn\\
 A_{22}^{\rm SMEFT}&=& A_{14}^{\rm SMEFT} (k_1 \leftrightarrow k_2,  k_3 \leftrightarrow k_4) \nn\\
 A_{23}^{\rm SMEFT}&=& A_{14}^{\rm SMEFT} (k_3 \leftrightarrow k_5)   \nn\\
 A_{24}^{\rm SMEFT}&=& A_{20}^{\rm SMEFT} (k_3 \leftrightarrow k_5)\nn\\
 A_{25}^{\rm SMEFT}&=& A_{14}^{\rm SMEFT} (k_1 \leftrightarrow k_2,  k_3 \leftrightarrow k_5)  \nn\\ 
A_{26}^{\rm SMEFT} &=& \frac{12 m_Z^2 }{v\Lambda ^2}c_{{\Phi D}}(\epsilon_1 \cdot \epsilon_2)
\eear

\section{Matching HEFT and SMEFT amplitudes}
\label{matching}
In order to clarify the way in which we proceed to perform the matching among HEFT and SMEFT,  we start first  analyzing  the simpler cases of $VV \to H$ and $HH \to HH$ amplitudes.  After this,  we will then solve the cases of interest here of $VV \to HH$ and $VV \to HHH$. 

The amplitude of single Higgs production from WBF,  $W(k_1)W (k_2) \to H (k_3) $,  using as input the physical parameters,  $m_W$,  $m_H$ and $v$,  is given in the HEFT and the SMEFT  respectively by:
\bear
A(WW \to H)^{\rm HEFT}&=&   \frac{2m_W^2}{v} a (\epsilon_1 \cdot \epsilon_2) \nn \\
&& - \frac{8m_W^2}{v^3}  a_{HWW} ((k_1 \cdot \epsilon_2) (k_2 \cdot \epsilon_1)- (k_1 \cdot k_2) (\epsilon_1 \cdot \epsilon_2))  \\
A(WW \to H)^{\rm SMEFT}&=&  \frac{2m_W^2}{v}  (\epsilon_1 \cdot \epsilon_2) \nn \\  
&&+ \frac{2v}{\Lambda^2} (m_W^2 (c_{\Phi \Box}- \frac{1}{4}c_{\Phi D})(\epsilon_1 \cdot \epsilon_2)) \nn \\
 &&+ 2 c_{\Phi W} ((k_1 \cdot \epsilon_2) (k_2 \cdot \epsilon_1)- (k_1 \cdot k_2) (\epsilon_1 \cdot \epsilon_2)))   
\eear 
Imposing the matching of these two amplitudes,  
\bear
A(WW \to H)^{\rm HEFT}&=& A(WW \to H)^{\rm SMEFT}
\label{matchingWWH}
\eear 
 requires the identification of all the Lorentz and gauge invariant structures involved.  Then,  if we write $a=1-\Delta a$,  the first terms in these two amplitudes coincide with the SM amplitude and this contribution cancels in both sides of the matching  \eqref{matchingWWH}.   Therefore,  solving the matching equation in terms of the EFT's coefficients leads to the following solution:
\bear
\Delta a &=& \frac{v^2}{\Lambda^2} (-c_{\Phi \Box}+ \frac{1}{4} c_{\Phi D}) \label{solutiondeltaa}  \\
a_{HWW}&=& -\frac{v^4}{2 m_W^2 \Lambda^2} c_{\Phi W} \label{solutionaHWW} 
\eear
It should be noticed  that the other coefficients entering in the Feynman rules of this $WWH$ vertex (see \appref{apHEFT}),  $a_{d2}$ and $a_{H \mV \mV}$ are not needed to match the HEFT and the SMEFT results at this order.  If instead, one goes beyond dim6 in the SMEFT,  i.e if one introduces dim8 operators in the SMEFT predictions,  then there could appear additional structures that might require the presence of these two HEFT coefficients to get the proper matching.  In consequence, at the established order in the two predictions these two coefficients 
$a_{d2}$ and $a_{H \mV \mV}$  can be ignored.  It should also be noticed that the same solution in 
Eqs. (\ref{solutiondeltaa}) and (\ref{solutionaHWW}) is obtained if the matching is instead applied to the decay amplitudes of the Higgs boson into two $W$'s (either real or virtual),  as expected.   

One can proceed similarly in the case of single Higgs production from the fusion of two $Z$ bosons,  $Z(k_1) Z(k_2) \to H(k_3)$,   now using as input parameters, 
$m_Z$, $m_H$ and $v$.  We also use here  the short notation given in \eqref{short}.  The corresponding amplitudes are:  
\bear
A(ZZ \to H)^{\rm HEFT}&=&   \frac{2m_Z^2}{v} a (\epsilon_1 \cdot \epsilon_2) \nn \\
&& - \frac{8m_Z^2}{v^3} 
( a_{HWW}c_W^4+ a_{HBB} s_W^4 + a_{H1} s_W^2 c_W^2) ((k_1 \cdot \epsilon_2) (k_2 \cdot \epsilon_1)- (k_1 \cdot k_2) (\epsilon_1 \cdot \epsilon_2))  \nn \\
&&- 8 \frac{m_Z^2}{v^3} s_W^2 m_Z^2 a_{H0}  (\epsilon_1 \cdot \epsilon_2)  \\
A(ZZ \to H)^{\rm SMEFT}&=&  \frac{2m_Z^2}{v}  (\epsilon_1 \cdot \epsilon_2) \nn \\  
&&+ \frac{2v}{\Lambda^2} (m_Z^2 (c_{\Phi \Box}+ \frac{1}{4}c_{\Phi D})(\epsilon_1 \cdot \epsilon_2)) \nn \\
 &&+ 2( c_{\Phi W} c_W^2 +  c_{\Phi B} s_W^2 +  c_{\Phi W B} s_W c_W) ((k_1 \cdot \epsilon_2) (k_2 \cdot \epsilon_1)- (k_1 \cdot k_2) (\epsilon_1 \cdot \epsilon_2)))   
\eear 
The matching equation in this case is:
\bear
A(ZZ \to H)^{\rm HEFT}&=& A(ZZ \to H)^{\rm SMEFT}
\label{matchingZZH}
\eear
Then,  the solution to this matching equation,  in addition to  the previous values for $\Delta a$ and $a_{HWW}$ in Eqs. (\ref{solutiondeltaa}) and (\ref{solutionaHWW}), is:
\bear
s_W^2 a_{H0}& =&-  \frac{v^4}{8 m_Z^2 \Lambda^2} c_{\Phi D} \label{solutionaH0}  \\
 s_W^2 a_{HBB} &=&- \frac{v^4}{2 m_Z^2 \Lambda^2}  c_{\Phi B}  \label{solutionaHBB} \\
 s_W c_W a_{H1} &=& - \frac{v^4}{2 m_Z^2 \Lambda^2}  c_{\Phi W B} \label{solutionaH1}
\eear

It is also illustrative to show the matching equation and its solution for the case of the $HH \to HH$ scattering amplitude.  Here we use the momentum assignments $H(k_1)H(k_2) \to H(k_3) H(k_4)$.  In this case, there are four contributing diagrams corresponding to:  contact c-channel,  s-channel,  t-channel  and u-channel. 
 We then write the amplitude by just using the subset of HEFT coefficients that are needed for the required matching.  In this case we have found that this subset is given by just 
 $\kappa_3$ and $\kappa_4$,  and the other HEFT coefficients can be ignored.   We then find the following contributions from the various channels,  correspondingly: 
\bear
A(HH \to HH)^{\rm HEFT} &=& A^{\rm HEFT}_c+A^{\rm HEFT}_s+A^{\rm HEFT}_t+A^{\rm HEFT}_u  \nn \\
&=&-\frac{3 \kappa_4 m_H^2}{v^2} -\frac{9 \kappa_3^2 m_H^4}{ (s-m_H^2) v^2} -\frac{9 \kappa_3^2 m_H^4}{ (t-m_H^2) v^2}
-\frac{9 \kappa_3^2 m_H^4}{ (u-m_H^2) v^2} 
\eear
\bear
A(HH \to HH)^{\rm SMEFT} &=& A^{\rm SMEFT}_c+A^{\rm SMEFT}_s+A^{\rm SMEFT}_t+A^{\rm SMEFT}_u  \nn \\
 &=&-\frac{3  m_H^2}{v^2} 
+ \frac{1}{\Lambda^2} \Big((-14 c_{\Phi \Box} +\frac{7}{2} c_{\Phi D}) m_H^2+ 36 c_\Phi v^2 \Big)  \nn \\
&&-\frac{9  m_H^4}{(s-m_H^2)v^2} \nn \\
&&+ \frac{3}{\Lambda^2}\frac{1}{(s-m_H^2)} \Big(
(-14 c_{\Phi \Box} +\frac{7}{2} c_{\Phi D}) m_H^4+(- 4 c_{\Phi \Box}+c_{\Phi D}) m_H^2 s + 12 c_\Phi m_H^2  v^2 \Big) \nn \\
&&-\frac{9  m_H^4}{(t-m_H^2)v^2} \nn \\
&&+ \frac{3}{\Lambda^2}\frac{1}{(t-m_H^2)} \Big(
(-14 c_{\Phi \Box} +\frac{7}{2} c_{\Phi D}) m_H^4+(- 4 c_{\Phi \Box}+c_{\Phi D}) m_H^2 t + 12 c_\Phi m_H^2  v^2 \Big) \nn \\
&&-\frac{9  m_H^4}{(u-m_H^2)v^2} \nn \\
&&+ \frac{3}{\Lambda^2}\frac{1}{(u-m_H^2)} \Big(
(-14 c_{\Phi \Box} +\frac{7}{2} c_{\Phi D}) m_H^4+(- 4 c_{\Phi \Box}+c_{\Phi D}) m_H^2 u + 12 c_\Phi m_H^2  v^2 \Big) \nn \\
\eear
 The matching equation in this case is:
\bear
A(HH \to HH)^{\rm HEFT} &=& A(HH \to HH)^{\rm SMEFT}  
\label{matchingHHHH}
\eear
To simplify this equation,  we first write in $A^{\rm HEFT}$ the $\kappa$' s as $\kappa_3 =1 - \Delta \kappa_3$ and  $\kappa_4 =1 - \Delta \kappa_4$ and  keep in the amplitude just the 
${\cal O}(\Delta^0)$  terms and  the linear terms in the $\Delta$'s,  of ${\cal O}(\Delta^1)$,  and we neglect the ${\cal O} (\Delta^2)$ terms.  Then,  the first contributions in $A^{\rm HEFT}$ from 
these ${\cal O}(\Delta^0)$  terms and the LO terms in $A^{\rm SMEFT}$ provide the SM amplitude which cancels in both sides of the matching equation.  Solving this matching equation then requires the identification of all Lorentz invariant structures involving $s$, $t$, $u$ and $m_H^2$ in this case.   
In this way,  the total predicted amplitude $A(s, m_H^2, \cos \theta)$ is an identical function of the energy $\sqrt{s}$,  mass $m_H^2$ and the relevant angle $\theta$ between $\vec{k}_3$ and $\vec{k}_1$,  in the CM frame for both theories.  Finally,  we find the following solution to the matching equation: 
\bear
\Delta \kappa_3 &=& \frac{2v^4}{m_H^2 \Lambda^2} c_\Phi + \frac{3v^2}{\Lambda^2} (-c_{\Phi \Box} + \frac{1}{4} c_{\Phi D}) \label{solutionk3} \\
\Delta \kappa_4 &=&  \frac{12v^4}{m_H^2 \Lambda^2} c_\Phi + \frac{50v^2}{3\Lambda^2} (-c_{\Phi \Box} + \frac{1}{4} c_{\Phi D})
\label{solutionk4}
\eear
At this point,  it is interesting to compare this solution in Eqs. (\ref{solutionk3}) and (\ref{solutionk4}) with the one obtained for $\Delta a$ in \eqref{solutiondeltaa}.  
We find the following  relations:
\bear 
\Delta \kappa_3 & =&  \frac{2v^4}{m_H^2 \Lambda^2} c_\Phi + 3 \Delta a  \label{relationk3} \\
\Delta \kappa_4 &=&   \frac{12v^4}{m_H^2 \Lambda^2} c_\Phi  +  \frac{50}{3} \Delta a \label{relationk4}
\eear
which in turn implies the following  interesting relation between $\Delta \kappa_3$,   $\Delta \kappa_4$ and $\Delta a$
\bear 
\Delta \kappa_4 &= & 6 \Delta \kappa_3 - \frac {4}{3} \Delta a 
\label{relationk3k4}
 \eear 
 Notice that when assuming $\Delta a=0$ we find a direct correlation between $\kappa_4$ and $\kappa_3$ within the HEFT given by $\Delta \kappa_4 =6 \Delta \kappa_3$ which is interesting to compare with the relation among the corresponding SM vertices $v V_{HHHH}^{\rm SM}=  V_{HHH}^{\rm SM}$.  This different relation in the HEFT as compared to the SM one could  provide different patterns of BSM versus SM signals in the  phenomenology of multiple Higgs production which deserves further studies. 
 
It is also remarkable that the above relation between $\Delta \kappa_3$ $\Delta \kappa_4$ and $\Delta a$ in \eqref{relationk3k4}  was also obtained in \cite{Salas-Bernardez:2022hqv} (in a different notation) by using a very different approach.  They worked in the pure scalar theory,   they did not consider  $c_{\Phi D}$,  and  performed  the matching of the two theories at the Lagrangian level,  by means of a series of scalar field transformations.

 Next we proceed to analyze the matching of our greatest interest,  i.e.,  the matching of the amplitudes for the cases  $VV \to HH$ and $VV \to HHH$.  For that purpose,  we will use our analytical results for these amplitudes within the HEFT and the SMEFT presented in the previous section.

In the case of $WW \to HH$,  the matching equation is:
\bear
A(WW \to HH)^{\rm HEFT} &=& A(WW \to HH)^{\rm SMEFT}  
\eear
where the results for  $A(WW \to HH)^{\rm HEFT}$ and $A(WW \to HH)^{\rm SMEFT}$,  using the momenta assignments $W^+(k_1), W^-(k_2) \to H (k_3) H(k_4)$,  are given in 
Eqs. (\ref{ampWWHH-HEFT}) and (\ref{ampWWHH-SMEFT}),  respectively.  Then we proceed as in the previous examples.  First we determine which is the subset of HEFT coefficients that are needed for the matching.  Next we identify all the terms in both amplitudes with the same Lorentz invariant structures that depend on $s$,  $t$,  $u$ and the masses involved,  
$m_W^2$ and $m_H^2$, in this case.  This,  in practice,  corresponds to identify  separately $A_s+A_c$,  $A_t$ and $A_u$.  Grouping  together the c-channel  and the s-channel in solving the matching equation can be understood because terms in the s-channel going as $s/(s-m_H^2)=1+m_H^2/(s-m_H^2)$ contain the '1' term that has the same structure than the c-channel.  Thus,  the full identification of the HEFT and SMEFT total amplitudes means that they are indeed the same analytical  function of the energy, masses and relevant angle,  $A(s, m_W^2, m_H^2, \cos\theta)$ (with $\theta$ in this case,  the angle between $\vec{k}_3$ and $\vec{k}_1$). 

For this case we find that the needed HEFT coefficients for the matching is the subset:  $a$,  $\kappa_3$,  $b$,  $a_{HWW}$ and $a_{HHWW}$.  The other coefficients either appear in combinations with these, therefore not producing new structures,  or else,  they provide new structures that  only appear  in the SMEFT at dim8.  This is,  for instance,  the case of  
$a_{dd \mV \mV 1}$ and $a_{dd \mV \mV 2}$ that,   as shown in \cite{Domenech:2022uud},  must be matched in particular with the dim8 SMEFT coefficients  $c_{\Phi^4}^{(i)}$ with $i=1,2,3$ 
(see this reference  for details).   One could aim for a more  complete  matching HEFT/SMEFT  including the full set of dim8 operators in the predictions of the SMEFT (for a very recent SMEFT computation of $VV \to HH$ including dim8 operators see \cite{Dedes:2025oda})  and by  repeating a similar exercise to the one we are doing here but including also all the additional needed HEFT NLO operators  that could match with the contributions to the amplitudes from the dim8 SMEFT operators.  But this is not our purpose here.  
Setting then the minimal subset of HEFT  coefficients to perform the matching  of HEFT-NLO with dim6-SMEFT then is equivalent to eliminate the other coefficients in the prediction of 
the scattering amplitude.  Specifically,  to solve the matching in this case we use the already simplified HEFT amplitude  in  \eqref{ampWWHH-HEFT} which contains just the needed coefficients for this matching with dim6-SMEFT.

Next we write the HEFT amplitudes in terms of 
 $\Delta a= 1-a$,  $\Delta b =1-b$,  $\Delta \kappa_3=1-\kappa_3 $,  $a_{HWW}$ and $a_{HHWW}$ and keep only the $ {\cal O}( \Delta^0)$ terms,  the linear terms in the $a_i$'s coefficients and the 
 $ {\cal O}( \Delta^1)$ terms.  The  first terms in the two amplitudes, HEFT and SMEFT,  provide the results of the SM amplitudes that then cancel  in both sides of the matching equation.  Then we find  the  solution to the matching equation given by the following predictions of the HEFT coefficients in terms of the SMEFT coefficients.  Firstly,   we get the same prediction for $\Delta a$  as in \eqref{solutiondeltaa},  the same prediction for $a_{HWW}$ as in \eqref{solutionaHWW},  and the same prediction for $\Delta \kappa_3$ as in \eqref{solutionk3}.  Secondly, we get in addition the following predictions for the remaining coefficients:
\bear
\Delta b &=&\frac{4 v^2}{\Lambda^2}  (-c_{\Phi \Box}+ \frac{1}{4} c_{\Phi D})  \label{solutiondeltab} \\
a_{HHWW}&=& -\frac{v^4}{4 m_W^2 \Lambda^2} c_{\Phi W} \label{solutionaHHWW}
\eear 
At this point, some comments are in order.   First,  we wish to note that,  in practice,  $\Delta a$ and $a_{HWW}$ are obtained from the matching using $A_t$ (similarly with $A_u$).  Once these two effective couplings are set, 
then  $\Delta b$  and $a_{HHWW}$ are obtained from the matching using $A_s+A_c$.  This means in particular that $\Delta b$ should not be extracted by identifying just the contact vertices, i.e.,   identifying $A_c$  in the two theories.  We wish to note that this apparently intuitive matching of contact vertices among the two EFTs  would have led to a wrong result for 
$\Delta b$.  This is equivalent to say that the identification between the HEFT and the SMEFT should not be done at the Feynman rules level (nor at the Lagrangian level) but at the amplitude level. 
It is the amplitude what is directly related with the observable quantity,  and doing the matching at the amplitude level provides the proper solution for the relations among HEFT and SMEFT coefficients which in turn have phenomenological implications.
Thus,  to solve properly for $\Delta b$ one must add the two channels, s-channel and c-channel.  Second,  we get the same solutions for $\Delta a$,  $\Delta \kappa_3$ and 
$\Delta a_{HWW}$ as in the previous examples.  The extra predictions now are $\Delta b$ and $\Delta a_{HHWW}$.   The above solutions from the matching between $A(WW \to HH)$  amplitudes  were already found previously in \cite{Domenech:2022uud}.   There is only one different result compared to this reference, which is that of $\Delta \kappa_3$.  The reason for this difference is that in \cite{Domenech:2022uud},  on one hand,  the chosen input parameters were not the same as here and, on the other hand,  the coefficient $c_\Phi$ was not included.  In \cite{Domenech:2022uud}, the input parameters were the Lagrangian parameters,  whereas here the input parameters,  as already said,  are the physical masses and $G_F$ (or equivalently $v$ extracted from $G_F$).  Finally,  we notice some interesting relations that are worth to comment.  We find: 
\bear
\Delta b &=& 4 \Delta a \label{relationdeltaadeltab} \\
2 a_{HHWW}&=&a_{HWW} \label{relationaHHWWaHWW} 
\eear
 The above relation between $\Delta a$ and $\Delta b$ was also obtained previously in the context of the pure scalar theory in \cite{Gomez-Ambrosio:2022qsi, Gomez-Ambrosio:2022why}. 
The previous result for the relations between  $\Delta a$ and $\Delta b$ and between  $a_{HHWW}$ and $a_{HHWW}$ should be compared with the corresponding relations among SM vertices,  $v V^{\rm SM}_{HHWW}= V^{\rm SM}_{HWW}$.  These different relations in the HEFT respect to the SM ones should provide again some hints in looking  for different patterns in BSM versus SM  predictions at double Higgs production.    

Next we analyze the $ZZ \to HH$ case.  The matching equation reads in this case,
\bear
A(ZZ \to HH)^{\rm HEFT} &=& A(ZZ \to HH)^{\rm SMEFT}  
\eear
We have found that for this case the needed subset of HEFT coefficients to perform the matching is given by:  $a$,  $\kappa_3$,  $b$,  $a_{HWW}$,  $a_{HHWW}$,  $a_{HBB}$,  
$a_{HHBB}$, $a_{H0}$, $a_{HH0}$,  $a_{H1}$ and $a_{HH1}$.  We then proceed with these coefficients,  as in the previous case,  using the linear approach in the $\Delta$'s and $a_i$'s again,  and canceling the SM-like contributions in both sides of the matching equation.  To shorten the computation,  we can then use the previous solutions for $\Delta a$ in  \eqref{solutiondeltaa},  $\Delta \kappa_3$ in \eqref{solutionk3},  $\Delta b$ in \eqref{solutiondeltab},   $a_{HWW}$, in
\eqref{solutionaHWW},  $a_{HHWW}$ in \eqref{solutionaHHWW},  $a_{HBB}$ in \eqref{solutionaHBB} , $a_{H0}$ in \eqref{solutionaH0}, and $a_{H1}$ in \eqref{solutionaH1}, and then solve for the remaining coefficients.  Thus,  we get the following additional relations among the HEFT and SMEFT coefficients from the  $ZZ \to HH$ case : 
\bear
s_W^2 a_{HHBB}&=&  - \frac{v^4}{4 m_Z^2 \Lambda^2}  c_{\Phi B}   \label{solutionaHHBB} \\
s_W^2 a_{HH0}& =& - \frac{5 v^4}{16 m_Z^2 \Lambda^2}   c_{\Phi D}  \label{solutionaHH0} \\
s_W c_W a_{HH1}& = &- \frac{v^4} {4 m_Z^2 \Lambda^2}  c_{\Phi W B} \label{solutionaHH1}
\eear
And, we find the following additional relations among the EFTs coefficients:
\bear 
2 a_{HHBB}&=&a_{HBB} \label{relationaHHBBaHBB}\\
2 a_{HH0}&=& 5 a_{H0} \label{relationaHH0aH0} \\
2 a_{HH1}&=&a_{H1} \label{relationaHH1aH1}
\eear
Finally, we analyze the matching in the triple Higgs boson production amplitudes. 
For the $WW \to HHH$ case,  the matching equation reads as follows:
\bear
A(WW \to HHH)^{\rm HEFT} &=& A(WW \to HHH)^{\rm SMEFT}  
\eear
where $A(WW \to HHH)^{\rm HEFT}$ and  $A(WW \to HHH)^{\rm SMEFT}$ are given in Eqs. (\ref{ampWWHHH-HEFT}) and (\ref{ampWWHHH-SMEFT}) respectively.   In this case,  the HEFT coefficients involved are those of the previous case $WW \to HH$, i.e., $a$,  $\kappa_3$,  $b$,  $a_{HWW}$, $a_{HHWW}$, and two more coefficients,  $\kappa_4$ and $c$.  

For solving the matching equation,  we proceed as in the previous cases.  First, we  write the HEFT amplitudes in terms of  $\Delta a= 1-a$,  $\Delta b =1-b$,  $\Delta \kappa_3=1-\kappa_3 $,  $\Delta \kappa_4=1-\kappa_4$,  $c$ and the involved $a_i$'s.  Then,  we  keep only the $ {\cal O}( \Delta^0)$ and $ {\cal O}( \Delta^1)$ terms,   the linear terms in $c$ and the linear terms in the $a_i$'s coefficients. The  first terms in the two amplitudes, HEFT and SMEFT,  provide the results of the SM amplitudes that then cancel  in both sides of the matching equation.  Next,  to simplify again the procedure,  we check first the previous solutions for $\Delta a$,  $\Delta b$,  $\Delta \kappa_3$,  $\Delta \kappa_4$,  $a_{HWW}$ and $a_{HHWW}$ by plugging them in the HEFT scattering amplitude.  Finally,  with the remaining contributions left from all the diagrams we then solve the matching equation for the new coefficient $c$.  To understand the behaviour of the different contributions  we group together in both cases,  HEFT and SMEFT,  the 26 diagrams into the 7 subsets,  S1-S7,  as explained in \secref{ampWWHHH-HEFT}.  This matching in practice can be solved by identifying the HEFT/SMEFT results for the sum of diagrams within these subsets for both EFTs.   Concretely,  we find that the analytical results for each diagram in S5 (all having two $W$ propagators) properly match the HEFT and SMEFT when using as input the  previously found coefficients.  Remember that the sum of diagrams in S5 is an invariant gauge quantity.  Next we consider the diagrams in  S2 and S3 that all have one $W$ propagator.  Remember that the sum of diagrams in S2+S3 is another  gauge invariant quantity.  Then we find that considering by pairs the sum of one diagram in S2 and other in S3  and using again the input values of the involved coefficients already computed, we get again a good HEFT/SMEFT matching.  Concretely,  these sums that fully match in HEFT and SMEFT are:  diag2+diag14,  diag3+diag19,   diag6+diag23,  diag7+diag17,  diag8+diag22,  diag11+diag25. 
Finally we consider the set of eight diagrams left, all of them having  zero $W$ propagators: diag1+diag4+diag5+diag9+diag12+diag13+diag16+diag26= S1+S4+S6+S7.  Remember that this sum is again another  gauge invariant quantity.  Then by plugging into this sum the values of all the previously coefficients found,  and identifying the two results in the HEFT and the SMEFT,  we solve finally for the value of the missing coefficient.  In this way we find the following solution for $c$:
\bear 
\label{solutionc}
c &=& -\frac{8 v^2}{3 \Lambda^2}  (-c_{\Phi \Box}+ \frac{1}{4} c_{\Phi D}) 
\eear
Notice that, similarly to what happened with the solution for $\Delta b$ in the $WW \to HH$ case, it would be a wrong procedure to compute $c$ by directly matching the amplitudes of the contact diagram,  diag26,  in the two EFTs.  Indeed,  since diag26  vanishes in the SMEFT,  this would have lead to a wrong prediction of a  vanishing $c$.  

Finally, we analize the $ZZ \to HHH$ case.  For this we follow exactly the same procedure as for the previous $WW \to HHH$ case.  Now the matching equation is: 
\bear
A(ZZ\to HHH)^{\rm HEFT} &=& A(ZZ \to HHH)^{\rm SMEFT}  
\eear
with $A(ZZ\to HHH)^{\rm HEFT}$ and  $A(ZZ \to HHH)^{\rm SMEFT}$ given in Eqs. (\ref{ampZZHHH-HEFT}) and (\ref{ampZZHHH-SMEFT}) respectively.  Now the HEFT coefficients involved are the previous ones in $WW\to HHH$,  i.e.,  $a$,  $b$,  $c$,  $\kappa_3$, $ \kappa_4$,  $a_{HWW}$,  and $a_{HHWW}$,  and in addition,  $a_{HBB}$,  $a_{HHBB}$,  $a_{H0}$, 
$a_{HH0}$,  $a_{HHH0}$,   $a_{H1}$ and $a_{HH1}$.  Following the same procedure as before,  we first plug in the values already obtained for the already known coefficients, and then solve for the remaining ones.  In this case , there is only one coefficient left to be determined by this new matching which is $a_{HHH0}$.  Thus,  we finally  get a perfect matching of the total amplitudes,  HEFT and SMEFT, by setting this new coefficient as follows:
\bear
\label{solutionaHHH0}
s_W^2 a_{HHH0}& =& - \frac{v^4}{4 m_Z^2 \Lambda^2}   c_{\Phi D}
\eear
With this last solution, the full set of coefficients for the matching of HEFT/SMEFT amplitudes is completed.

\section{High energy behaviour of multiple Higgs production from $V_LV_L$}
Finally, in this section we study the high energy  behaviour of  the scattering amplitudes for double and triple Higgs production in the particular case of longitudinally polarized EW gauge bosons.  We also consider the single Higgs  production case for comparison.
We focus here on the amplitudes $A(W_LW_L \to HH)$ and $A(W_LW_L \to HHH)$.  We wish to explore the high energy behaviour  of these amplitudes by means of an analytical expansion in powers of a small parameter written generically as
$m_{\rm EW}/\sqrt{s}$,  where $\sqrt{s}$ is the energy of the process and 
$m_{\rm EW}$ refers to the EW masses involved and $v$.  In the case of $W_LW_L$ fusion this implies 
$\sqrt{s} \gg m_{W}, m_{H}, v$. 
The interest of this study is twofold: on the one hand, it serves us to test the prediction done in the literature using the Equivalence Theorem,  where these amplitudes are  approximated by the corresponding amplitudes with external massless GBs and where these  later are computed within the pure scalar theory \cite{Contino:2010mh,Contino:2013gna,Delgado:2023ynh}.  On the other hand,  this computation will also serve us to show the different 
energy behaviour that the HEFT and the SMEFT amplitudes have.  This being,  in turn,  very relevant for their respective tests at colliders.

For this exercise we start with our HEFT amplitudes in  \secref{amplitudes} and particularize them  to the case where the $W^{\pm}$ polarization vectors,  $\epsilon_1$ and $\epsilon_2$ are set to the longitudinal ones,  $\epsilon_1^L$ and $\epsilon_2^L$ respectively.  
 Since we search  only for the leading contributions to the amplitudes found in \secref{amplitudes} in this large $s$ expansion,  which turn out to be linear in $s$,  i.e.  of ${\cal O} (s)$,  we use some starting simplifying assumptions.  First, we consider only the  terms  in the various diagrams that only involve the LO-HEFT coefficients.  The other terms can be shown to be subleading i.e.  of ${\cal O}(s^{0})$.   We have already mentioned explicitly those LO-terms in \secref{amplitudes}.  Second, we approximate the longitudinal polarization vectors respectively by $\epsilon_1^L \simeq k_1/m_W$ and $\epsilon_2^L \simeq k_2/m_W$.  Third,  we consider all the Lorentz invariant quantities involving products of two external four-momentum.  Generically, products of two different four-momenta are quantities of order $s$, i.e.,  $k_i . k_j = {\cal O}(s^1)$, for $i \neq j$,  whereas for two equal four-momenta  they are $k_i^2=m_i^2={\cal O}(s^0)$.   Then,  using the same notation as in \secref{amplitudes},  we get the following results for the leading terms of all the diagrams involved in the high energy limit.
 
\subsection{Case  $HH$}

In the double Higgs production case,  for the first terms in this high energy expansion,  we get:
\bear
A_{c}^{\rm HEFT}&\simeq &2  \, b \, v^{-2} \, m_W^2 \, (\epsilon_1 \cdot \epsilon_2) ={\cal O}(s^1)   \nn\\
A_{s}^{\rm HEFT}&\simeq &\frac{6 \, a \, \kappa_3 \, v^{-2} \, m_H^2 \, m_W^2 \,}{q_{12}^2-m_H^2 }(\epsilon_1 \cdot \epsilon_2)={\cal O}(s^0) \nn\\
A_{t}^{\rm HEFT}&\simeq &\frac{4 \, a^2 \, v^{-2} \, m_W^2}{k_{13}^2-m_W^2 }  \left( - (\epsilon_1 \cdot k_{13}) \, (\epsilon_2 \cdot k_{13}) \right) = {\cal O}(s^1) \nn\\
A_{u}^{\rm HEFT}& \simeq & \frac{4 \, a^2 \, v^{-2} \, m_W^2}{k_{14}^2-m_W^2 }  \left( - (\epsilon_1 \cdot k_{14}) \, (\epsilon_2 \cdot k_{14}) \right) = {\cal O}(s^1) 
\eear
Therefore,   by adding them and considering $k_1+k_2=k_3+k_4$,  we  get from the sum of $A_c$, $A_t$ and $A_u$ the total result of ${\cal O}(s)$,  where  the angular dependence in $A_t$ and $A_u$ cancels,  leading to:
\bear
A^{\rm HEFT}(W_L W_L \to HH)& \simeq & -\frac{1}{v^2}(a^2-b) s
\label{ampWWHH-largeE}
\eear

Similarly one can conclude on the high energy behaviour result of single Higgs production  from WBF of two longitudinal $W$'s. From the first equation in \secref{matching} we find,  straightforwardly:
\bear
A^{\rm HEFT}(W_L W_L \to H) \simeq \frac{1}{v} a s
\label{ampWWH-largeE}
\eear

\subsection{Case  $HHH$}

In the triple Higgs production case we get,  for the 7 reference diagrams: 
\bear
A_1^{\rm HEFT}& \simeq &6 m_W^2 m_H^2 v ^{-3} \frac{ 1}{q_{45}^2-m_H^2} (b \, \kappa_3) (\epsilon_1 \cdot \epsilon_2)= {\cal O}(s^0) \nn \\
A_2^{\rm HEFT}& \simeq &4 m_W^4  v^{-3}  \frac{ 1}{k_{13}^2-m_W^2} (a\, b)
 \left(-m_W^{-2} (\epsilon_1 \cdot k_{13})(\epsilon_2 \cdot k_{13}) \right)={\cal O}(s^1) \nn \\
 A_4^{\rm HEFT}& \simeq & 6m_W^4m_H^2v^{-3}\frac{(a \kappa_4)}{q_{12}^2-m_H^2}(\epsilon_1\cdot \epsilon_2)= {\cal O}(s^0)\nn \\
 A_5^{\rm HEFT}&\simeq &  \frac{18m_W^2m_H^4 v^{-3}}{(q_{12}^2-m_H^2)(q_{34}^2-m_H^2)}(a \kappa_3^2)(\epsilon_1\cdot \epsilon_2) ={\cal O}(s^{-1}) \nn\\
 A_{14}^{\rm HEFT}&\simeq &12m_W^4m_H^2v^{-3}\frac{(a^2\kappa_3)}{(k_{13}^2-m_W^2)(q_{45}^2-m_H^2)}\left(-m_W^{-2}(\epsilon_1\cdot k_{13})(\epsilon_2\cdot k_{13})\right) =
 {\cal O}(s^0) \nn\\
 A_{20}^{\rm HEFT} &\simeq & 8 m_W^6 v^{-3}  \frac{1}{(k_{13}^2-m_W^2)(k_{25}^2-m_W^2)} (a^3) 
 \left( +  m_W^{-4} (\epsilon_1 \cdot k_{13}) (\epsilon_2 \cdot k_{25}) (k_{25} \cdot k_{13}) \right) = {\cal O}(s^1) \nn \\
 A_{26}^{\rm HEFT}&\simeq &6  m_W^2 v^{-3}  \,(c) \, (\epsilon_1 \cdot \epsilon_2)={\cal O}(s^1)
\eear
The other diagrams can be  obtained from the above reference ones by the proper change in momenta,  as defined in \eqref{ampWWHHH-HEFT}. 
Then,  by grouping the 26 diagrams by subsets,  and using $k_1+k_2=k_3+k_4+k_5$,  we get:
\bear
S1 & =& {\cal O}(s^0)  \,\,; \,\, S2 \simeq - \frac{4}{v^3} (ab) s \,\, ; \, \, S3  =  {\cal O}(s^0) \,\,; \, \, S4 =  {\cal O}(s^0)  \nn \\
S5 & \simeq & \frac{4}{v^3} (a^3) s \,\,; \,\, S6  =  {\cal O}(s^0)  \,\, ; \, \, S7  \simeq  \frac{3}{v^3} (c) s 
\eear
Finally,  adding the leading subsets S2,  S5 and S7  we  get the resulting total amplitude of  ${\cal O}(s)$:
\bear
A^{\rm HEFT}(W_L W_L \to HHH)& \simeq & \frac{3}{v^3} \big(c + \frac{4}{3} a(a^2-b) \big) s
\label{ampWWHHH-largeE}
\eear 
It is worth noticing  that the above results in Eqs.  (\ref{ampWWHH-largeE}) and (\ref{ampWWHHH-largeE}) coincide with the approximate result found in  \cite{Contino:2013gna,Delgado:2023ynh} where they used the Equivalence Theorem and computed with external massless GBs in the pure scalar theory.  

Finally, to end this section it is illustrative to show the particular high-energy results  of these longitudinal WBF amplitudes in two cases: 1) when comparing HEFT and SM , and 2) when comparing HEFT and SMEFT.  With this purpose in mind,  we first write the results in Eqs. \ref{ampWWH-largeE},  \ref{ampWWHH-largeE} ,  and \ref{ampWWHHH-largeE}  in terms of $\Delta a$,  $\Delta b$ and $c$.  This leads to the generic BSM predictions at large energies within the LO-HEFT: 
\bear
A^{\rm HEFT}(W_L W_L \to H)& \simeq & \frac{s}{v} (1- \Delta a) \nn \\
A^{\rm HEFT}(W_L W_L \to HH)& \simeq & \frac{s}{v^2} (2 \Delta a - \Delta b)  \nn \\
A^{\rm HEFT}(W_L W_L \to HHH)& \simeq & \frac{s}{v^3} 3 (c-\frac{8}{3} \Delta a + \frac{4}{3} \Delta b)  
 \label{HEFT-largeE}
\eear
When setting the HEFT parameters to the SM values,  i.e.  for $\Delta a=0$,  $\Delta b= 0$ and $c=0$,  we get the well known SM results:
\bear
A^{\rm SM}(W_L W_L \to H)& \simeq &  \frac{s}{v} (1) \nn \\
A^{\rm SM}(W_L W_L \to HH)& \simeq &  \frac{s}{v^2} (0)  + {\cal O}(s^0) \nn \\
A^{\rm SM}(W_L W_L \to HHH)& \simeq & \frac{s}{v^3} (0)  + {\cal O}(s^0)   
 \label{SM-largeE}
\eear
which shows that at high energies compared to the involved boson masses, the double and triple Higgs production in the SM are suppressed with respect to the BSM prediction,  since the SM amplitudes  do not grow with the CM squared energy. 
We can also use the above generic HEFT results to get the SMEFT results.  This is done by the simple substitution of the previous values found for the $\Delta a$,  $\Delta b$ and $c$ from the matching,  i.e.  by the values in Eqs. (\ref{solutiondeltaa}), (\ref{solutiondeltab}) and (\ref{solutionc}) respectively.  Then,  we get for the SMEFT the following results:
\bear
A^{\rm SMEFT}(W_L W_L \to H)& \simeq & \frac{s}{v}  \big (1- \frac{v^2}{\Lambda^2} (-c_{\Phi \Box}+ \frac{1}{4} c_{\Phi D}) \big) \nn \\
A^{\rm SMEFT}(W_L W_L \to HH)& \simeq & \frac{s}{v^2} \big (-2  \frac{v^2}{\Lambda^2} (-c_{\Phi \Box}+ \frac{1}{4} c_{\Phi D})
\big) \nn \\
A^{\rm SMEFT}(W_L W_L \to HHH)& \simeq & \frac{s}{v^3} (0)  + {\cal O}(s^0) 
 \label{SMEFT-largeE}
\eear
The interesting  results above show that triple Higgs production from WBF of two longitudinal  $W$'s  vanishes within the dim6-SMEFT at 
${\cal O}(s)$, as it happens in the SM case.  This is in remarkable contrast with the  result of the LO-HEFT that grows linearly with $s$.  This may have important implications for the  BSM Higgs searches at future colliders.  For a recent discussion on some phenomenological implications of these analytical results of  the HEFT versus SMEFT within the context of multiple Higgs production at both lepton a hadron colliders,  see for instance Ref. \cite{Anisha:2024ryj} where the full bosonic sector (scalar and gauge) is considered.   This has also been explored in the simplified scenario of the pure scalar theory in  Refs. \cite{Contino:2010mh,Contino:2013gna,Delgado:2023ynh}.

\section{Conclusions}
\label{conclu}
\begin{table}[!h]
    \centering
    \begin{tabular}{|c|cc|}
        \hline
        \multicolumn{3}{|c|}{Matching dim6-SMEFT to HEFT} \\
        \hline\hline 
        LO-HEFT & \multicolumn{2}{|c|}{NLO-HEFT} \\
        \hline
        $\Delta a = \frac{v^2}{\Lambda^2}\left(-c_{\Phi\Box}+\frac{1}{4}c_{\Phi D}\right)$ & $s_W^2 a_{H0} = -\frac{v^4}{8m_Z^2\Lambda^2}c_{\Phi D}$ & $a_{HWW} = -\frac{v^4}{2m_W^2\Lambda^2}c_{\Phi W}$ \\
        
        $\Delta b = \frac{4v^2}{\Lambda^2}\left(-c_{\Phi\Box}+\frac{1}{4}c_{\Phi D}\right)$ & $s_W^2 a_{HH0} = -\frac{5v^4}{16m_Z^2\Lambda^2}c_{\Phi D}$ & $a_{HHWW} = -\frac{v^4}{4m_W^2\Lambda^2}c_{\Phi W}$ \\

        $c = -\frac{8v^2}{3\Lambda^2}\left(-c_{\Phi\Box}+\frac{1}{4}c_{\Phi D}\right)$ & $s_W^2 a_{HHH0} = -\frac{v^4}{4m_Z^2\Lambda^2}c_{\Phi D}$ & $s_W^2 a_{HBB} = -\frac{v^4}{2m_Z^2\Lambda^2}c_{\Phi B}$ \\

        $\Delta \kappa_3 = \frac{2v^4}{m_H^2\Lambda^2}c_\Phi + \frac{3v^2}{\Lambda^2}\left(-c_{\Phi\Box}+\frac{1}{4}c_{\Phi D}\right)$ & & $s_W^2 a_{HHBB} = -\frac{v^4}{4m_Z^2\Lambda^2}c_{\Phi B}$ \\

        $\Delta \kappa_4 = \frac{12v^4}{m_H^2\Lambda^2}c_\Phi + \frac{50v^2}{3\Lambda^2}\left(-c_{\Phi\Box}+\frac{1}{4}c_{\Phi D}\right)$ & & $s_W c_W a_{H1} = -\frac{v^4}{2m_Z^2\Lambda^2}c_{\Phi WB}$ \\

        & & $s_W c_W a_{HH1} = -\frac{v^4}{4m_Z^2\Lambda^2}c_{\Phi WB}$ \\[8pt]

        \hline

    \end{tabular}
    \caption{ \label{tab:matching-table}Summary of the results found from the matching of HEFT/SMEFT amplitudes}
    \end{table}
 In this paper we have explored the issue of matching HEFT and SMEFT  by means of the identification of scattering amplitudes in both theories.  For this purpose we have chosen a set of 7 specific processes that have direct connection with experimental data.  These selected processes are:  single,  double and triple Higgs production from WBF,  $VV \to H$, $VV \to HH$,  $ VV \to HHH$ with $VV=WW, ZZ$  and the elastic scattering   $HH \to HH$.  For the solving of the matching equations we have truncated the predictions of the HEFT at NLO, i.e, including effective operators up to chiral dimension 4,  and the predictions of the SMEFT at NLO, i.e.  including effective operators up to canonical dim6.   
The solutions to the HEFT/SMEFT  matching equations,  when imposed all together and  for all the helicity amplitudes involved,  lead to fix in total   14 HEFT coefficients in terms of the 6 SMEFT coefficients defining the dim6-SMEFT.  The  set of solutions  found in this work  are summarized in \tabref{tab:matching-table}.  These contain the solutions for the 5 LO-HEFT coefficients,  $\Delta a$, $\Delta b $, $c$,  $\Delta \kappa_3$ and $\Delta \kappa_4$   and for the 9 NLO-HEFT coefficients: $a_{H0}$,  $a_{HH0}$,  $a_{HHH0}$,  $a_{HWW}$,   $a_{HHWW}$,  $a_{HBB}$,   $a_{HHBB}$, $a_{H1}$ and $a_{HH1}$.    As it is well known,  the matching of HEFT/SMEFT coefficients occurs in crossed orders of these EFTs.  The LO-HEFT coefficients are given in terms of the three dim6 Wilson coefficients referring to the doublet $\Phi$,  i.e.,   $c_{\Phi\Box}$,  $c_{\Phi D}$ and $c_{\Phi}$.  The NLO-HEFT coefficients $a_{H0}$,  $a_{HH0}$ and $a_{HHH0}$ are given in terms of just one SMEFT coefficient,  $c_{\Phi D}$.  The remaining NLO-HEFT coefficients in the last column of the table are given in terms of the SMEFT coefficients involving weak  gauge bosons,  $W$ and $B$.  It is also interesting to notice that when considering the weak isospin limit (also referred to as imposing custodial symmetry) given by 
$m_Z=m_W$ (or,  equivalently,  $g'=0$, or $s_W=0$)  we get an important  reduction in the relevant coefficients among these later involving weak gauge bosons in the third column.  These coefficients  are $a_{HWW}$ and $a_{HHWW}$ which are both given in terms of one SMEFT coefficient $c_{\Phi W}$. 
    
It is also worth commenting that in order to fix other NLO-HEFT coefficients of \eqref{eq-L4},  the matching with the SMEFT should go beyond dim6. In particular, focusing in the most relevant operators for BSM Higgs signals at colliders in the TeV energy regime, which are those containing four derivatives, the coefficients in front of those operators, $a_{dd{\cal VV}1}$ and 
$a_{dd{\cal VV}2}$, have been proven in~\cite{Domenech:2022uud} to match with the SMEFT coefficients in front of the three dim8 operators containing also four derivatives, $C_{\varphi^4D^4}^{(1),(2),(3)}$ in the notation of~\cite{Dedes:2023zws}. Concretely the matching of HEFT/SMEFT amplitudes for $WW \to HH$ in~\cite{Domenech:2022uud} finds the following relations among these coefficients:\\
$a_{dd{\cal VV}1}=\frac{v^4}{4 \Lambda^4}(C_{\varphi^4D^4}^{(1)}+C_{\varphi^4D^4}^{(2)})$ and $a_{dd{\cal VV}2}=\frac{v^4}{4 \Lambda^4}C_{\varphi^4D^4}^{(3)}$.

Besides the previous predictions for the HEFT coefficients in terms of the SMEFT ones,  we have also found a set of interesting relations among EFT coefficients.  Concretely,  when  assuming  the SMEFT to be the low energy theory for BSM Higgs Physics,  the following  summary in \eqref{relations-conclu}  of relations are found among the corresponding HEFT coefficients:
\bear
\Delta b &=& 4 \Delta a \,\, ; \, \, c= -\frac{8}{3} \Delta a =  - \frac{2}{3} \Delta b \,\,; \, \, 
\Delta \kappa_4 =  6 \Delta \kappa_3 - \frac {4}{3} \Delta a  \nn\\
 a_{HHWW}&=& \frac{1}{2} a_{HWW} \, \,; \, \, a_{HHBB}= \frac{1}{2} a_{HBB} \, \, ; \, \,a_{HHH0} = \frac{4}{5} a_{HH0}= 2 a_{H0}   
\label{relations-conclu}
\eear
These relations may have important impact for the search of BSM Higgs signals  at future colliders.  This is because they may lead 
to correlations among specific observables that can be measured in the future experiments.  In particular,  we have investigated here one of those observables that will be relevant for the future colliders exploring the multiple Higgs production from WBF.  Concretely,  by an explicit analytical computation with the full bosonic particle content (scalar and gauge),  we have explored the behaviour of the double and triple Higgs production from longitudinal WBF at high energies compared with the boson masses involved.  We have demonstrated analytically in \eqref{HEFT-largeE} that in the generic HEFT, both the double and the triple amplitudes grow with energy as ${\cal O}(s)$.  In contrast,  for the SMEFT  we have found in \eqref{SMEFT-largeE} that the double production also grows with energy  as ${\cal O}(s)$,  but  the triple Higgs production does not grow with $s$,  leading to a vanishing result to ${\cal O}(s)$ (as in the SM case).  We then get,  analytically,  the suppression of triple Higgs production in SMEFT in contrast to HEFT,   where this production will be enhanced (respect to the SM prediction) at large energies compared to the boson masses.  This implies that the search for triple Higgs production and the comparison with double Higgs production at colliders will provide a very powerful tool to disentangle the two EFTs.  Most probably, this is nothing else than a potential signal disentangling the singlet/doublet Higgs nature in both approaches HEFT/SMEFT which in our opinion deserves further study.


\section*{Acknowledgments}
D.D. and M.J.H. acknowledge partial financial support by the Spanish Research Agency (Agencia Estatal de Investigación) through the grant  with reference number PID2022-137127NB-I00 
funded by MCIN/AEI/10.13039/501100011033/ FEDER, UE.  They also acknowledge financial support from the  grant IFT Centro de Excelencia Severo Ochoa with reference number CEX2020-001007-S
 funded by MCIN/AEI/10.13039/501100011033, from the previous AEI project PID2019-108892RB-I00 funded by MCIN/AEI/10.13039/501100011033, and from the European Union’s Horizon 2020 research and innovation programme under the Marie Sklodowska-Curie grant agreement No 860881-HIDDeN.  
 The work of R.A.M. is supported by CONICET and ANPCyT under project PICT-2021-00374.  ASB akwnowledges the support of grants PID2023-148162NB-C21 and PID2022-137003NB-I00 from Spanish MCIN/AEI/10.13039/501100011033/ and EU FEDER.
The work of D.D. is also supported by the Spanish Ministry of Science and Innovation via an FPU grant No FPU22/03485. 
\newpage



\section*{Appendices}
\appendix
Here we summarize the relevant Feynman rules for double and triple Higgs production from WBF in the SM,  HEFT-NLO and SMEFT-NLO.
In all the vertices we use the convention of all momenta ingoing and the following specific assignments of momenta and Lorentz indices: 
\begin{eqnarray}
 H(p_1)H(p_2)H(p_3) & , & H(p_1)H(p_2)H(p_3)H(p_4) \nn\\
H(p_1)W^{+\mu_2}(p_2) W^{-\mu_3}(p_3) & , &  H(p_1)H(p_2) W^{+\mu_3}(p_3) W^{-\mu_4}(p_4) \nn\\
 H(p_1)Z^{\mu_2}(p_2) Z^{\mu_3}(p_3)  & , &  H(p_1)H(p_2)  Z^{\mu_3}(p_3) Z^{\mu_4}(p_4) \nn\\
 H(p_1)H(p_2)H(p_3) W^{+\mu_4}(p_4) W^{-\mu_5}(p_5)   & , &  H(p_1)H(p_2)H(p_3) Z^{\mu_4}(p_4) Z^{\mu_5}(p_5) \nn 
\end{eqnarray}
In the Feynman rules presented in these Appendices  we use the short notation,  omitting indices:
\bear
\Gamma_{HHH} & \equiv &  \Gamma_{H_1 H_2 H_3} \equiv \Gamma (H(p_1), H(p_2), H(p_3)) \nn \\
\Gamma_{HHHH} & \equiv &  \Gamma_{H_1 H_2 H_3 H_4} \equiv \Gamma (H(p_1), H(p_2), H(p_3), H(p_4)) \nn \\
\Gamma_{HVV}  &\equiv &   \Gamma ^{\mu_2 \mu_3}_{H_1 V_2 V_3} \equiv  \Gamma (H(p_1), V^{\mu_2 }(p_2), V^{\mu_3}(p_3)) \nn \\
\Gamma_{HHVV}  &\equiv &   \Gamma ^{\mu_3 \mu_4}_{H_1 H_2 V_3 V_4} \equiv  \Gamma (H(p_1), H(p_2), V^{\mu_3 }(p_3), V^{\mu_4}(p_4)) \nn \\
\Gamma_{HHHVV}  &\equiv &   \Gamma ^{\mu_4 \mu_5}_{H_1 H_2 H_3 V_4 V_5} \equiv  \Gamma (H(p_1), H(p_2), H(p_3), V^{\mu_4 }(p_4), V^{\mu_5}(p_5)) \nn 
\eear 
We also use a short notation for the frequently appearing $W$ and $Z$ mass combinations:
\begin{equation}
c_W^2 \equiv \frac{m_W^2}{m_Z^2} \, , \, s_W^2 \equiv  \frac{m_Z^2-m_W^2}{m_Z^2} \, , \,   s_W c_W \equiv  \frac{m_W \sqrt{m_Z^2-m_W^2}}{m_Z^2} \nn
\end{equation}

\section{Feynman Rules of SM}
\label{apSM}
For a clear comparison,  we remind first the SM Feynman rules for these vertices when expressed in terms of the physical input parameters,  i.e.,  
the masses $m_H$,  $m_W$,  $m_Z$ and the vev related to the Fermi constant,  $v=(\frac{1}{\sqrt{2} G_F})^{1/2} =246$ GeV.
\bear
i\Gamma_{HHH}^{\rm SM} &=&-3i\frac{\mh^2}{\vev}  \nn\\
i\Gamma_{HHHH}^{\rm SM}  &=& -3i\frac{\mh^2}{\vev^2}  \nn\\
i\Gamma_{HWW}^{\rm SM}  &=& 2i\frac{m_W^2}{\vev}g^{\mu_2 \mu_3}  \nn\\
i\Gamma_{HHWW}^{\rm SM}  &=& 2i\frac{m_W^2}{\vev^2}g^{\mu_3\mu_4}  \nn\\
i\Gamma_{HZZ}^{\rm SM}  &=&2i\frac{m_Z^2}{v} g^{\mu_2\mu_3}  \nn\\
i\Gamma_{HHZZ}^{\rm SM}  &=& 2i\frac{m_Z^2}{v^2}g^{\mu_3\mu_4}  \nn\\
i\Gamma_{HHHWW}^{\rm SM}  &=&  0 \nn\\
i\Gamma_{HHHZZ}^{\rm SM}  &=&  0 
\label{FR-SM}
\eear
It is inmediate to deduce from the above expressions the well known relations between SM pairs of vertices: 
\bear
v V_{HHHH}^{\rm SM}& = & V_{HHH}^{\rm SM} \nn \\
v V_{HHWW}^{\rm SM} &= & V_{HWW}^{\rm SM} \nn  \\
v V_{HHZZ}^{\rm SM} & = & V_{HZZ}^{\rm SM} 
\eear 
These relations are a consequence of $H$ being a component of a $SU(2)$ doublet within the SM.
Within the HEFT and SMEFT,  however,  these relations do no hold,  as it is shown through this work.  In the HEFT case these vertices are not correlated,  even at LO,  since $H$ is a singlet.   In the SMEFT-NLO,  they are related but in a different way  than in the SM case due to the contributions from the dim6 operators.  

\section{Feynman Rules of HEFT-NLO}
\label{apHEFT}
In the following we display the Feynman rules of the HEFT-NLO that we also express
in terms of the physical parameters,  the masses $m_H$,  $m_W$,  $m_Z$ and the vev $v$  related to the Fermi constant.
\begin{eqnarray}
\textcolor{red}{\bullet} \,\, i\Gamma_{HHH}^{\rm HEFT} &=&-3i \textcolor{red}{\kappa_3} \frac{\mh^2}{\vev}  \nn\\
&&+\frac{i}{\vev^3}\left(a_{Hdd}\mh^2+a_{ddW}\mw^2+a_{ddZ}\mz^2\right)(p_1^2+p_2^2+p_3^2)+{\cal O}(p^4) \nn\\
\textcolor{cyan}{\bullet} \,\,i\Gamma_{HHHH}^{\rm HEFT}  &=& -3i \textcolor{red}{\kappa_4} \frac{\mh^2}{\vev^2}  \nn\\
&&+\frac{2i}{\vev^4}\left(a_{HHdd}\mh^2+a_{HddW}\mw^2+a_{HddZ}\mz^2\right)(p_1^2+p_2^2+p_3^2+p_4^2) +{\cal O}(p^4) \nn\\
\textcolor{yellow}{\bullet}\,\,i\Gamma_{HWW}^{\rm HEFT}  &=& 2i \textcolor{red}{a}\frac{m_W^2}{\vev}g^{\mu_2 \mu_3}  \nn\\
&&-\frac{2im_W^2}{\vev^3}\left(a_{d2}(p_1^{\mu_2}p_2^{\mu_3}+p_1^{\mu_3}p_3^{\mu_2}+g^{\mu_2\mu_3}p_1^2) \right.  \nn\\
&&\left. -a_{H\mV\mV}\mh^2g^{\mu_2\mu_3} +4 \textcolor{red}{a_{HWW}}(p_2^{\mu_3}p_3^{\mu_2}-g^{\mu_2\mu_3}p_2\cdot p_3)\right)  \nn\\
\textcolor{violet}{\bullet} \,\,i\Gamma_{HHWW}^{\rm HEFT}  &=& 2i \textcolor{red}{b} \frac{m_W^2}{\vev^2}g^{\mu_3\mu_4}  \nn\\
&&-2i\frac{m_W^2}{v^4}\left(a_{Hd2}((p_1+p_2)^{\mu_3}p_3^{\mu_4}+(p_1+p_2)^{\mu_4}p_4^{\mu_3}+g^{\mu_3\mu_4}(p_1+p_2)^2) \right.  \nn\\
&&\left. -a_{HH\mV\mV}\mh^2g^{\mu_3\mu_4} +8 \textcolor{red}{a_{HHWW}}(p_3^{\mu_4}p_4^{\mu_3}-g^{\mu_3\mu_4}p_3\cdot p_4)\right.  \nn\\
&&\left. +2a_{dd\mV\mV1}(p_1^{\mu_3}p_2^{\mu_4}+p_1^{\mu_4}p_2^{\mu_3}) +4a_{dd\mV\mV2}g^{\mu_3\mu_4}p_1\cdot p_2\right)  \nn\\
\textcolor{yellow}{\bullet}\,\,i\Gamma_{HZZ}^{\rm HEFT}  &=&2i \textcolor{red}{a} \frac{m_Z^2}{v} g^{\mu_2\mu_3}  \nn\\
&&-2i\frac{m_Z^2}{v^3} \left((a_{d2} c_W^2-a_{d1} s_W^2)(p_1^{\mu_2}p_2^{\mu_3}+p_1^{\mu_3}p_3^{\mu_2}+g^{\mu_2\mu_3}p_1^2) \right.  \nn\\
&&\left. -a_{H\mV\mV}\mh^2g^{\mu_2\mu_3} +4(\textcolor{red}{a_{HWW}} c_W^4+
\textcolor{red}{a_{HBB}} s_W^4+ \textcolor{red}{a_{H1}} s_W^2 c_W^2)(p_2^{\mu_3}p_3^{\mu_2}-g^{\mu_2\mu_3}p_2\cdot p_3)   \right. \nn\\
&& \left. +4 \textcolor{red} {a_{H0}} s_W^2 m_Z^2 g^{\mu_2\mu_3} \right) \nn\\
\textcolor{violet}{\bullet} \,\,i\Gamma_{HHZZ}^{\rm HEFT}  &=& 2i \textcolor{red}{b}\frac{m_Z^2}{v^2}g^{\mu_3\mu_4}  \nn\\
&&-2i\frac{m_Z^2}{v^4}\left((a_{Hd2} c_W^2-a_{Hd1} s_W^2)((p_1+p_2)^{\mu_3}p_3^{\mu_4}+(p_1+p_2)^{\mu_4}p_4^{\mu_3}+g^{\mu_3\mu_4}(p_1+p_2)^2) \right.  \nn\\
&&\left. -a_{HH\mV\mV}\mh^2g^{\mu_3\mu_4} +8(\textcolor{red}{a_{HHWW}}c_W^4+
\textcolor{red}{a_{HHBB}} s_W^4+\textcolor{red}{a_{HH1}} s_W^2 c_W^2)(p_3^{\mu_4}p_4^{\mu_3}-g^{\mu_3\mu_4}p_3\cdot p_4)\right.  \nn\\
&&\left. +2a_{dd\mV\mV1}(p_1^{\mu_3}p_2^{\mu_4}+p_1^{\mu_4}p_2^{\mu_3}) +4a_{dd\mV\mV2}g^{\mu_3\mu_4}p_1\cdot p_2\right)  \nn\\
&&+8 \textcolor{red}{a_{HH0}} s_W^2 m_Z^2 g^{\mu_3\mu_4} ) \nn \\
\,\textcolor{green}{\bullet} \,\,i\Gamma_{HHHWW}^{\rm HEFT}  &=& 6i \textcolor{red}{c} \frac{m_W^2}{v^3} g^{\mu_4 \mu_5}+ {\cal O}(m^2p^2)+ {\cal O}(p^4) \nn\\
\,\textcolor{green}{\bullet} \,\,i\Gamma_{HHHZZ}^{\rm HEFT}  &=& 6i \textcolor{red}{c} \frac{m_Z^2}{v^3} g^{\mu_4 \mu_5} -48i \frac{m_Z^4}{v^5} \textcolor{red}{a_{HHH0}} s_W^2 g^{\mu_4 \mu_5} +  {\cal O}(m^2p^2)+ {\cal O}(p^4)
\label{FR-HEFT}
\end{eqnarray}
First,  notice that the set of SM Feynman rules summarized in Appendix A can be reached from these HEFT Feynman rules above by fixing $a=b=\kappa_3=\kappa_4=1$,  $c=0$ and setting to zero all the NLO coefficients,  $a_i=0$.
Second, we have used a different color code  on the left  of the vertices above to appreciate more clearly the place in the Feynman diagrams of the scattering amplitudes where they appear in the present work.  Third, notice also that we have specified in red color the particular HEFT coefficients that appear in the matching equations discussed in this work.  Finally, notice 
that in the above expressions of $\Gamma_{HHH}^{\rm HEFT}$ and $\Gamma_{HHHH}^{\rm HEFT}$ we have not specified the ${\cal O}(p^4)$ terms,  which go as four powers of momentum since they are not needed for the present work.  We have not included either the ${\cal O}(p^4)$  contributions  in the contact interactions,  $\Gamma_{HHHVV}^{\rm HEFT}$,  for the same reason.


\section{Feynman Rules of  SMEFT-NLO}
\label{apSMEFT}

Here we collect the Feynman Rules of SMEFT-NLO.  We use again as input parameters the physical masses  $m_H$,  $m_W$, $m_Z$ and $v$ related to the Fermi constant. 
\begin{eqnarray}
\textcolor{red}{\bullet} \,\, i\Gamma_{HHH}^{\rm SMEFT}&=&- 3i\frac{m_H^2}{v} +i  \frac{1}{\Lambda^2} \big( 6 \, c_\Phi v^3 +  v (\frac{1}{4} c_{\phi D} - c_{\Phi \Box})
\big( 3 m_H^2 
+2 (p_1^2+p_2^2+p_3^2) \big) \big) \nn \\
\textcolor{cyan}{\bullet} \,\,i\Gamma_{HHHH}^{\rm SMEFT}&=&-3i \frac{m_H^2}{v^2} + i \frac{1}{\Lambda^2} \big( 
36 \, c_\Phi v^2 +  (\frac{1}{4} c_{\Phi D}-c_{\Phi\Box}) (6m_H^2 +2(p_1^2+p_2^2+p_3^2+p_4^2)) \big) \nn \\
\textcolor{yellow}{\bullet} \,\,i\Gamma_{HWW}^{\rm SMEFT}&=&2i \frac{m_W^2}{v} g^{\mu_2 \mu_3}+ i \frac{v}{\Lambda^2} \big(-2 m_W^2 ( \frac{1}{4} c_{\Phi D}- \, c_{\Phi \Box}  ) \, g^{\mu_2 \mu_3} 
+ 4 \, c_{\Phi W}  \big( p_2^{\mu_3} p_3^{\mu_2} - g^{\mu_2 \mu_3} \, p_2 \cdot p_3 \big) 
\big) \nn \\
\textcolor{violet}{\bullet} \,\,i\Gamma_{HHWW}^{\rm SMEFT}&=&2i\frac{ m_W^2 }{v^2} g^{\mu_3 \mu_4}+i \frac{1}{\Lambda^2} \big(-4 m_W^2 (\frac{1}{4} \, c_{\Phi D}-  c_{\Phi \Box} ) \, g^{\mu_3 \mu_4} 
+ 4  \, c_{\Phi W} \big( p_3^{\mu_4} p_4^{\mu_3} - g^{\mu_3 \mu_4} \, p_3 \cdot p_4 \big) 
\big) \nn \\
\textcolor{yellow}{\bullet}\,\,i \Gamma_{HZZ}^{\rm SMEFT}&=&2i \frac{m_Z^2}{v} g^{\mu_2 \mu_3}
+ i \frac{v}{\Lambda^2} \big(2 m_Z^2 (\frac{1}{4} c_{\Phi D}+c_{\Phi \Box}) \,g^{\mu_2 \mu_3} \nn \\
&&\quad + (4 \, c_{\Phi W}  c_W^2 
+ 4 \, c_{\Phi B}  s_W^2
+ 4 \, c_{\Phi WB} s_W c_W)
\big( p_2^{\mu_3} p_3^{\mu_2} - g^{\mu_2 \mu_3} \, p_2 \cdot p_3 \big) 
\big) \nn \\
\textcolor{violet}{\bullet} \,\,i \Gamma_{HHZZ}^{\rm SMEFT}&=&2i\frac{m_Z^2}{v^2} g^{\mu_3 \mu_4}
+ i \frac{1}{\Lambda^2} \big( 
4 m_Z^2 ( c_{\Phi \Box} +  \, c_{\Phi D}) \, g^{\mu_3 \mu_4} \nn \\
&&\quad + (4 \, c_{\Phi W} c_W^2 + 4 \, c_{\Phi B} s_W^2 
+4 \, c_{\Phi WB} s_W c_W )
\big( p_3^{\mu_4} p_4^{\mu_3} - g^{\mu_3 \mu_4} \, p_3 \cdot p_4 \big) 
\big)  \nn \\
\,\textcolor{green}{\bullet} \,\,i \Gamma_{HHHWW}^{\rm SMEFT}&=&0+0+\mathcal{O}(\frac{1}{\Lambda^{4}}) \nn \\
\textcolor{green}{\bullet} \,\,i \Gamma_{HHHZZ}^{\rm SMEFT}&=& 0+i\frac{1}{\Lambda ^2 v} 12 m_Z^2 \, c_{\Phi D} \, g^{\mu _4 \mu _5}
\label{FR-SMEFT}
\end{eqnarray}
Notice that the first terms in these SMEFT vertices coincide with  the SM values given in Appendix A.  The second terms in these vertices are of $ \mathcal{O}(\frac{1}{\Lambda^{2}})$.   Then,  it is interesting to remark that,  in the SMEFT,  the $HHHWW$ contact vertex does not  appear at this order,  $ \mathcal{O}(\frac{1}{\Lambda^{2}})$.  In contrast,  the $HHHZZ$ contact vertex it does appear.  None of these two contact vertices appear in the SM case,  so they are generic BSM Higgs interactions.  Notice also that we have used the same color code for the various vertices as those in Appendix B.  This helps to identify in which place of the Feynman diagrams they are contributing to the scattering amplitudes studied here. 

\newpage

\bibliography{DHMS-PRDv2}

\end{document}